\documentclass{aa}

\usepackage{graphicx}
\usepackage{txfonts}

\usepackage{txfonts}
\usepackage{wasysym}
\usepackage{enumitem,amssymb}
\usepackage{pifont}
\usepackage{array}
\usepackage[flushleft]{threeparttable}
\usepackage{siunitx}

\usepackage{lscape}
\usepackage{placeins}

\usepackage{amsmath}
\usepackage{amssymb}
\usepackage{tabularx}
\usepackage{comment}
\usepackage{svg}

\usepackage{xcolor}

\usepackage[flushleft]{threeparttable}

\usepackage{natbib,twoopt}
\usepackage[breaklinks=true]{hyperref} 
\bibpunct{(}{)}{;}{a}{}{,}             
\makeatletter
  \newcommandtwoopt{\citeads}[3][][]{\href{http://adsabs.harvard.edu/abs/#3}%
    {\def\hyper@linkstart##1##2{}%
     \let\hyper@linkend\@empty\citealp[#1][#2]{#3}}}
  \newcommandtwoopt{\citepads}[3][][]{\href{http://adsabs.harvard.edu/abs/#3}%
    {\def\hyper@linkstart##1##2{}%
     \let\hyper@linkend\@empty\citep[#1][#2]{#3}}}
  \newcommandtwoopt{\citetads}[3][][]{\href{http://adsabs.harvard.edu/abs/#3}%
    {\def\hyper@linkstart##1##2{}%
     \let\hyper@linkend\@empty\citet[#1][#2]{#3}}}
  \newcommandtwoopt{\citeyearads}[3][][]%
    {\href{http://adsabs.harvard.edu/abs/#3}
    {\def\hyper@linkstart##1##2{}%
     \let\hyper@linkend\@empty\citeyear[#1][#2]{#3}}}
\makeatother

\usepackage{color}
\hypersetup{colorlinks=true,linkcolor=blue,citecolor=blue,urlcolor=blue}

\usepackage[breaklinks=true]{hyperref}

\def\kms{\,km\,s$^{-1}$}

\def\Mearth{\hbox{$\mathrm{M}_{\oplus}$}}

\def\degr{\hbox{$^\circ$}}

\begin{document}

   \title{The star HIP~41378 potentially misaligned with its cohort of long-period planets
   \thanks{The radial velocity data for all instruments are available in electronic form at the CDS via anonymous ftp to cdsarc.u-strasbg.fr (130.79.128.5) or via \url{http://cdsweb.u-strasbg.fr/cgi-bin/qcat?J/A+A/}.}}

\author{S.~Grouffal\inst{1}
          \and
          A.~Santerne\inst{1}
          \and
          V.~Bourrier\inst{2}
          \and
          V.~Kunovac\inst{3,4}
          \and
          C.~Dressing\inst{5}
          \and 
          B.~Akinsanmi\inst{2}
          \and
          C.~Armstrong\inst{3,4}
          \and
          S.~Baliwal\inst{6,7}
          \and
          O.~Balsalobre-Ruza\inst{8}
          \and
          S.C.C.~Barros\inst{9,10}
          \and
          D.~Bayliss\inst{3,4}
          \and
          I.J.M.~Crossfield\inst{11}
          \and
          O.~Demangeon\inst{9}
          \and
          X.~Dumusque\inst{2}
          \and
          S.~Giacalone\inst{12}
          \and 
          C.K.~Harada\inst{5}
          \and
          H.~Isaacson\inst{5}
          \and
          H.~Kellermann\inst{13}
          \and
          J.~Lillo-Box\inst{8}
          \and
          J.Llama\inst{14}
          \and
          A.~Mortier\inst{15}
          \and
          E.~Palle\inst{16,17}
          \and
          A.S.~Rajpurohit\inst{6}
          \and
          M.~Rice\inst{18}
          \and
          N.C.~Santos\inst{9,10}
          \and
          J.V.~Seidel\inst{19,20}
          \and
          R.~Sharma\inst{6}
          \and
          S.G.~Sousa\inst{9}
          \and
          L.~Thomas\inst{13,21}
          \and
          E.V.~Turtelboom\inst{5}
          \and
          S.~Udry\inst{2}
          \and
          P.J.~Wheatley\inst{3,4}
          }

   \institute{Aix Marseille Univ, CNRS, CNES, Institut Origines, LAM, Marseille, France
         \and
            Observatoire de Genève, Université de Genève, Chemin Pegasi 51, 1290 Versoix, Switzerland
        \and
        Centre for Exoplanets and Habitability, University of Warwick, Coventry, CV4 7AL, UK
        \and
        Department of Physics, University of Warwick, Coventry, CV4 7AL, UK
        \and
        Department of Astronomy, 501 Campbell Hall \#3411, University of California, Berkeley, CA 94720, USA
        \and
        Astronomy \& Astrophysics Division, Physical Research Laboratory, Ahmedabad, India 
        \and
        Indian Institute of Technology, Gandhinagar, India 
        \and
        Centro de Astrobiolog\'ia (CAB, CSIC-INTA), Depto. de Astrof\'isica, ESAC campus, 28692, Villanueva de la Ca\~nada (Madrid), Spain
        \and
        Instituto de Astrof\'isica e Ci\^encias do Espa\c{c}o, Universidade do Porto, CAUP, Rua das Estrelas, 4150-762 Porto, Portugal
        \and
         Departamento de F\'isica e Astronomia, Faculdade de Ci\^encias, Universidade do Porto, Rua do Campo Alegre, 4169-007 Porto, Portugal
         \and
         Department of Physics and Astronomy, University of Kansas, Lawrence, KS, USA
         \and
         Department of Astronomy, California Institute of Technology, Pasadena, CA 91125, USA
         \and
         University Observatory Munich, Faculty of Physics, Ludwig-Maximilians-Universit\"at München, Scheinerstr. 1, 81679 Munich, Germany
         \and
         Lowell Observatory, 1400 W Mars Hill Rd. Flagstaff. AZ
         \and
         School of Physics \& Astronomy, University of Birmingham, Edgbaston, Birmingham, B15 2TT, UK
         \and
         Instituto de Astrof\'{i}sica de Canarias (IAC), 38205 La Laguna, Tenerife, Spain
         \and
         Departamento de Astrof\'isica, Universidad de La Laguna (ULL), E-38206 La Laguna, Tenerife, Spain
         \and
         Department of Astronomy, Yale University, New Haven, CT 06511, USA
         \and
         European Southern Observatory, Santiago, Chile
         \and
         Laboratoire Lagrange, Observatoire de la Côte d’Azur, CNRS, Université  Côte d’Azur, Nice, France
         \and
         Max-Planck Institute for Extraterrestrial Physics, Giessenbachstrasse 1, D-85748 Garching, Germany
             }

   \date{Received 12 May 2025; accepted 31 July 2025}

  \abstract
   {The obliquity between the stellar spin axis and the planetary orbit, detected via the Rossiter-McLaughlin (RM) effect, is a tracer of the formation history of planetary systems. While obliquity measurements have been extensively applied to hot Jupiters and short-period planets, they remain rare for cold and long-period planets due to observational challenges, particularly their long transit durations. We report the detection of the RM effect for the 19-hour-long transit of HIP 41378 f, a temperate giant planet on a 542-day orbit, observed through a worldwide spectroscopic campaign. We measure a slight projected obliquity of $21\pm8$\degr\ and a significant 3D spin-orbit angle of $52 \pm 6$\hbox{$^\circ$, based on the measurement of the stellar rotation period.} HIP 41378 f is part of a 5-transiting planetary system with planets close to mean motion resonances. The observed misalignment likely reflects a primordial tilt of the stellar spin axis relative to the protoplanetary disk, rather than dynamical interactions. HIP~41378~f is the first non-eccentric long-period (P>$100$ days) planet observed with the RM effect, opening new constraints on planetary formation theories. This observation should motivate the exploration of planetary obliquities across a longer range of orbital distances through international collaboration.}


   \maketitle
%

\section{Introduction}

Exoplanetary systems are providing us with important constraints in understanding the physical processes of planet formation, migration, and evolution. Among the information we might obtain from these distant worlds, the stellar obliquity, defined as the angle between a star's spin axis and a planet's orbital axis, is an excellent tracer of their history. This obliquity is measured mainly through the Rossiter-McLaughlin (RM) effect \citep{Holt1893,Rossiter1924, McLaughlin1924}, which is seen as a Doppler anomaly caused by a transiting planet obscuring the rotating stellar surface. Over the last decades, the RM effect has been used to understand the migration processes of hot Jupiters, for which many have been reported to be misaligned with their host star \citep[see][and references therein]{2022PASP..134h2001A}. Although the exact mechanisms behind these misalignments are still open, one possible scenario is orbital perturbation that could increase the obliquity \citep{Dawson2018}. 

Expanding the sample of measured obliquities towards long-period planets allows us to probe systems that underwent different formation and migration mechanisms \citep{Rice2021}. In particular, they have not been significantly sculpted by stellar tidal realignment \citep[e.g.][]{Davies19, Rice2022}. Multi-planetary systems also provide us with crucial tests for primordial disk tilting. Initial studies by \citet{Albrecht2013} revealed that multi-planetary systems tend to be aligned; however, multi-planet systems with significant misalignment have also been observed, suggesting primordial misalignment \citep{Hjorth2021}.

In this paper, we report the spin-orbit orientation of the planet HIP 41378 f, with an orbital period of 542 days, which is the longest-period planet observed through the RM effect to date. HIP 41378 f orbits a bright (V=8.93), late F-type star with an effective temperature of $T_{eff} = 6290$ K. It is the outermost transiting planet in the system discovered by the K2 mission \citep{Vanderburg2016}. Previous RM measurements for planet d (period of 278 days) revealed orbital misalignment \citep{Grouffal2022}. However, the signal was near the detection limits due to the planet’s small size $R_p = 3.54 \pm 0.06$ \hbox{$\mathrm{R}_{\oplus}$}. In contrast, planet f, with a radius three times larger than planet d, offers a clear detection to measure the system’s spin-orbit misalignment.

Observing long-period planets via the RM effect is difficult due to their infrequent and long-duration transits. For instance, HIP 41378 f has an exceptionally long transit event of 19 hours, making continuous observation challenging from a single ground-based observatory. To overcome this, we coordinated a global spectroscopic campaign involving nine optical high-precision instruments. These observations, for the first time, characterise orbital misalignment in a long-period system and provide critical insights into the origins of stellar inclinations in dynamically complex architectures. 

\section{Observations and data reduction}
\label{Obs and data reduction}

\begin{figure}[h!]
   \centering
   \includegraphics[width=1.0\columnwidth]{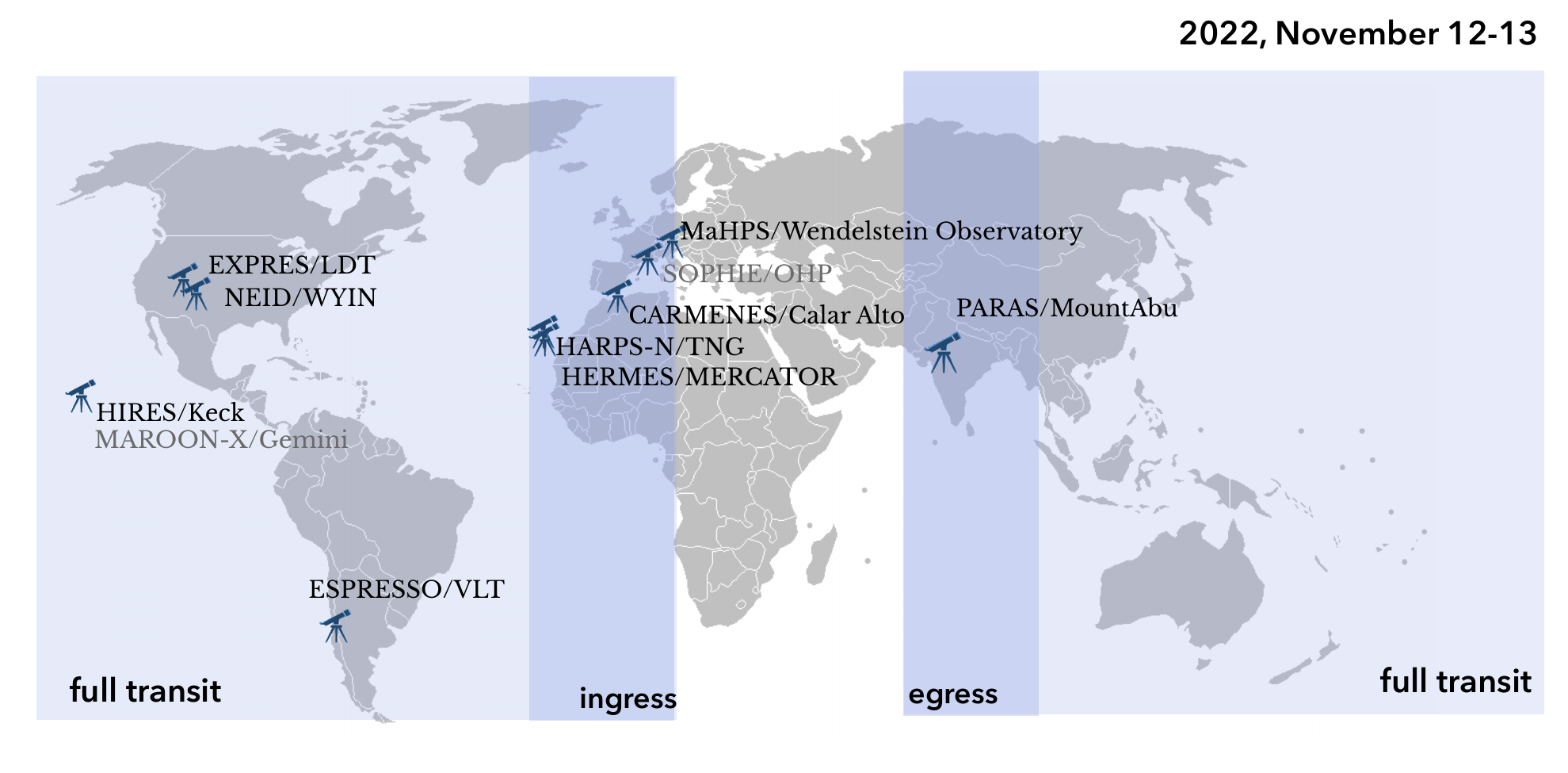}
      \caption{Maps of the spectrographs involved in observing HIP 41378 f's transit during the night of 2022, November 12-13. The regions of the world where the transit was visible are highlighted in blue. Ingress and egress are in darker blue. The spectrographs that could not provide data due to bad weather conditions or technical problems are written in grey, whereas the others are in black.}
         \label{fig_instruments}
   \end{figure}

Although the 19-hour transit exceeds the length of a single night from a location on Earth, the trajectory of the planet can be observed almost in its entirety by leveraging the rotation of Earth through a coordinated campaign. Figure \ref{fig_instruments} shows the locations of all spectrographs involved in the monitoring of the transit. Ingress was seen from HARPS-N, CARMENES, HERMES, and MaHPS. SOPHIE was also scheduled to observe, but was hindered by poor weather conditions. The middle of the transit was observed with ESPRESSO, EXPRES, NEID and HIRES. Finally, the end of the egress was observed with PARAS. Through this multi-site strategy, we successfully covered 70$\%$ of the transit. The remaining gaps correspond to longitudes in which no high-precision radial-velocity (RV) instruments are available.\\

\paragraph{HARPS-N} The ingress of the transit of HIP 41378 f was observed with the high-resolution (R$\sim$115000), optical (383-693 nm) HARPS-N spectrograph \citep{2012SPIE.8446E..1VC} at the 3.6-m Telescopio Nazionale Galileo (TNG) at the Roque de Los Muchachos Observatory on the island of La Palma, Spain, as part of the programme ID OPT22B\_15. Those observations were performed near the ingress time (between UTC 02:05 and 06:41). More spectra were collected the night following the transit event to extend the baseline. In total, we obtained 102 spectra of HIP 41378 with an exposure time of 300~s, leading to a Signal-to-noise (SNR) per pixel up to 70 at 550 nm. The observations were performed with the sky in the second fibre to monitor the sky background that would be important to remove for atmospheric study. The instrument is stable enough to allow precise RV constraints over one night. The spectra were reduced using the online pipeline, version 3.7, available at the observatory. This pipeline is based on the cross-correlation technique \citep{Baranne1996} to compute an averaged line-profile and determine the instantaneous RV. This pipeline is routinely used in HARPS-N observations and publications.\\

\paragraph{CARMENES} The CARMENES spectrograph \citep[R$\sim$94600; ][]{2018SPIE10702E..0WQ} at the 3.5m telescope of the Calar Alto Observatory in Spain also monitored the transit ingress contemporaneously with HARPS-N. The observations were obtained as part of the programme ID OPT22B\_15. A total of 18 spectra were collected over the transit night with an exposure time of 900s, leading to a SNR up to 150 in the spectral order 50. A Fabry-Perot lamp illuminated the second fibre to monitor any instrumental drift within the transit night. Because the near-infrared arm of CARMENES was not stable enough and polluted by tellurics, we only used the visible arm to derive the RV (550-950 nm). The basic reduction of the spectra was carried out at the observatory with the CARACAL pipeline (\citealt{zechmeister14, bauer15}) version 2.20. RVs were subsequently extracted with the \texttt{SHAQ} pipeline, specifically developed for the KOBE experiment \citep{lillo-box22}. This cross-correlation-based \citep{Baranne1996} pipeline employs publicly available binary masks from the ESPRESSO instrument pipeline (\citealt{Pepe21}). \texttt{SHAQ} has demonstrated compatibility with results obtained through template-matching algorithms (\citealt{balsalobre25}).\\

\paragraph{MaHPS} We observed the ingress of the transit of HIP~41378~f with the MaHPS spectrograph installed on the 2.1m telescope at the Wendelstein Observatory. MaHPS consists of there-fed, high-resolution ($R \sim 60,000$), optical ($385 - 885$ nm) echelle spectrograph FOCES \citep{Pfeiffer1998} and a Menlo Systems Laser Frequency Comb (LFC) as a wavelength calibration source. The observations lasted from 23:30 UT until 05:45 UT and covered a baseline before the transit and the ingress. The exposure time was 900~s for each observation, and we used simultaneous exposures of the LFC to correct for drifts during the observation. In total, we collected 22 spectra. At the beginning of the observation, the target was below 30 $^\circ$ elevation. We, therefore, excluded the first five spectra from the analysis. The data was reduced with the \texttt{GAMSE} pipeline \citep{Wang2016}, and RVs were extracted using \texttt{MARMOT} \citep{kellermann2021commissioning} following the analysis in \citet{Thomas2025}.\\

\paragraph{ESPRESSO} The transit was also observed with the high-precision (R$\sim$140000), visible (380-788 nm) ESPRESSO spectrograph \citep{Pepe21}, mounted on the 8.2-m ESO-VLT (VLT) at Paranal Observatory in Chile, as part of the programme ID 104.20UB.004. Baseline observations were not secured as the transit occurred during a technical night near the middle of a 2-week interferometric campaign using the four VLTs. A total of 31 spectra (between UTC 06:25 and 09:12) were secured with an exposure time of 300~s, leading to a SNR per pixel at the level of 50, at 450~nm. The RVs were extracted using the ESPRESSO data-reduction software in its version 3.2.5. \\ 

\paragraph{NEID} On UT 13 November 2022, we observed HIP 41378 using the high-resolution ($R \sim 110,000$) NEID optical spectrograph (380-930 nm) on the 3.5-m WIYN telescope at Kitt Peak National Observatory in AZ, USA \citep{schwab2016}. We collected 25 spectra of HIP 41378 for $\sim$4 hours covering the second quarter of the transit of planet f (between UTC 08:33 and 11:54). Exposure times were fixed to 600 seconds, and the data were reduced with version 1.2.0 of the NEID Data Reduction Pipeline. This version includes an update to the calibration files, which were required due to the thermal cycling of the NEID spectrometer after the Contreras fire. RVs and their uncertainties were obtained with the Cross-Correlation Function (CCF) method, resulting in measurement uncertainties of 3-4 m$\,\mathrm{s}^{-1}$. In addition, we obtained six spectra respectively on UT nights 12 and 14 November 2022 to establish the out-of-transit baseline.\\ 

\paragraph{EXPRES} We observed HIP 41378 during a transit of planet f on UT night 13 November 2022 using the EXPRES spectrograph \citep{jurgenson2016} mounted on the 4.3-m Lowell Discovery Telescope \citep{levine2012,levine2016}, located near Flagstaff, AZ, USA. EXPRES is a stabilised high-resolution ($R\sim 137,000$) optical (390-780 nm) spectrograph with a photon noise precision of about 33 cm$\,\mathrm{s}^{-1}$. We took 16 exposures over a duration of $\sim4.2$ hours covering the second quarter of the 19-hour duration transit (between UTC 09:01 and 13:15). Exposure times ranged from 600-900 seconds, resulting in a SNR between 43-69 at 550 nm. A LFC and Th-Ar lamp was used for wavelength calibration throughout the observations. We computed CCFs from the extracted 1D spectra using a F9 line mask. RVs and their measurement uncertainties were derived from the centre of a Gaussian fit to the CCF, resulting in a RV precision of 2-4 m$\,\mathrm{s}^{-1}$. We also obtained three exposures in sequence on UT night 12 November 2022 and 7 exposures on UT night 14 November 2022 to establish the out-of-transit baseline.\\ 

\paragraph{HIRES} We acquired RV observations of HIP~41378 on 12 November 2022, 13 November 2022, and 14 November 2022 using the High Resolution Echelle Spectrometer (HIRES, \citep{Vogt1994}) on the Keck I telescope at W. M. Keck Observatory. HIRES is an echelle spectrometer with a resolving power of $\sim$67,000, depending on the slit used, and a wavelength range spanning from 374 to 970 nm. We extracted RVs from the spectroscopic data (reduced using the standard procedure described in \citet{Howard2010}), using the iodine cell method outlined in \citet{Butler1996}. Observations were taken using a 14" x  0.861" slit with exposure times determined by an internal exposure meter. The typical exposure lasted 4-5 minutes, resulting in a spectrum that reached a signal-to-noise ratio per resolution element of ~200 near 550nm. The observations on 13 November 2022 were acquired directly before, during, and after the ingress of the transit of HIP~41378~f and covered approximately half of the transit (between UTC 11:42 and 15:40). The observations on 12 November 2022 and 14 November 2022 were acquired to help establish the out-of-transit RV trend.\\

\paragraph{PARAS} The end of the transit egress was visible by the PARAS spectrograph. PARAS (now known as PARAS-1) is a fibre-fed, high-resolution (R $\approx$ 67000), echelle spectrograph that is pressure and temperature-stabilised, operating within the wavelength range of 380 to 690 nm to facilitate precise RV measurements \citep{Chakraborty2014}. It is mounted on the PRL 1.2-m telescope located at the Gurushikhar observatory, Mt. Abu, India. The spectrograph has shown 1-3 m s$^{-1}$ long-term RV stability on bright RV standard stars \citep{Chakraborty2014, Sharma2021, Chakraborty2018}. A total of six spectra of HIP~41378 were acquired using PARAS. These spectra were collected on November 12 and 13, 2022, utilising the simultaneous wavelength calibration mode with a Uranium-Argon (UAr) hollow cathode lamp (HCL), as outlined in \citet{Sharma2021}. The exposure durations were set at 1200 seconds and 1800 seconds for the different spectra, leading to an SNR per pixel ranging from 13 to 19 at the blaze wavelength of 550 nm. Data reduction was performed using an automated, custom-designed pipeline developed in IDL \citep{Chakraborty2014}, based on the algorithms described by \citet{Piskunov2002}. The resulting extracted spectra from this reduction process were subsequently used for RV measurements. RVs were obtained by cross-correlating the stellar spectra with a numerical template mask representative of the same spectral type \citep{Baranne1996, Pepe2002}. The errors provided here are based on the fitting errors of the CCF and the photon noise, computed similarly to the methods presented in \citet{Chaturvedi2016, Chaturvedi2018}.\\

\paragraph{HERMES} 23 spectra were obtained with the high-resolution (R$\sim$85000) fibre-fed spectrograph HERMES mounted on the Flemish 1.2m Mercator telescope on the Roque de los Muchachos observatory in Spain \citep{raskin11}. This instrument is equipped with a science-grade pipeline which returns the response corrected, wavelength-calibrated 1D full spectrum from about 390 to 900nm on the barycentric velocity frame. Ten spectra were collected at ingress time (between UTC 02:52 and 06:40) and 12 spectra after the transit. \\

\paragraph{K2 short-cadence photometry} HIP 41378 was initially discovered and subsequently re-observed by the Kepler space telescope during the K2 extended mission, with data collected in Campaign 5 (2015; \citealt{Vanderburg2016}) and Campaign 18 (2018; \citealt{2019AJ....157..185B, 2019AJ....157...19B}). On both occasions, a single transit of HIP 41378 f was recorded. The Campaign 18 observations were obtained in short-cadence mode, offering improved temporal resolution. We incorporated this high-cadence K2 photometry into our analysis, alongside the RM effect detection, to refine the planetary parameters and constrain the system’s true spin-orbit obliquity. The short-cadence data from Campaign 18 were processed using the K2SFF pipeline \citep{Vanderburg2014}, which corrects for the telescope’s pointing drifts. To preserve the transit shape and depth while mitigating systematics, we applied a self-flat-fielding approach that explicitly accounts for the presence of the transit. Remaining instrumental or stellar variability was modelled and removed using a cubic spline with knots spaced every 10 days, resulting in a cleaner, transit-preserving light curve suitable for joint modelling with the RM observations.

\section{Analysis}
\label{Analysis}

\subsection{Stellar rotation}\label{subsec42}
An accurate understanding of HIP~41378’s stellar rotation is critical for determining the true obliquity of the system. The short-cadence K2 C18 data allowed \citet{Lund2019} to provide stellar parameters for HIP~41378 through asteroseismic characterisation. Combining the star's radius and rotation period gives us an estimate of its equatorial rotational velocity.

The rotational modulation of HIP 41378 is clearly visible in the K2 light curve. We used the short-cadence Campaign 18 light curve, reduced with the EVEREST pipeline, to estimate the stellar rotation period. The transits of planets b, c, d, and f were masked using the ephemerides and transit durations from \citet{2019AJ....157...19B}. An initial estimate with a Lomb-Scargle periodogram \citep{Zechmeister2009} gives a rotation period of $P_{rot} = 6.4$ days (see figure \ref{fig_rotation_period}). To confirm this result, we also applied a Bayesian Generalised Lomb-Scargle periodogram \citep{Mortier2015}, which revealed a single prominent peak at approximately 6.4 days. Finally, an autocorrelation function (ACF) analysis following \citet{McQuillan2013} indicated a similar period of roughly 6.7 days (see Figure~\ref{fig_rotation_period}), consistent with the previous estimates. 
To further refine the stellar rotation period, we applied quasi-periodic Gaussian Process (GP) modelling using the \texttt{celerite2} package, which implements a kernel composed of two simple harmonic oscillators centred at $P_{rot}$ and $P_{rot}/2$ \citep{2018MNRAS.474.2094A}. A uniform prior was adopted for the rotation period over the range 4–20 days. The GP model yields an inferred rotation period of  $P_{rot}=6.8 \pm 0.4$~days.

In addition to photometry, the radial-velocity (RV) analysis of ESPRESSO data (Grouffal et al. in prep.) revealed a stellar rotation period of $P_{\mathrm{rot}} = 7.6 \pm 1.4$ days, as inferred from the $\log{R'_{HK}}$ index, along with a signal in the periodogram near 8 days. To be as conservative as possible, all available data suggest that the stellar rotation period lies between 6 and 9 days. We therefore adopted a uniform prior between 6 and 9 days for the rotation period of the star in the following analysis.

\subsection{Planetary system parameters}\label{subsec43}
We jointly modelled the RM observation and the photometric data from K2 using a combination of the \textsf{starry} \citep{2019AJ....157...64L} and \textsf{exoplanet} \citep{exoplanet:joss,
exoplanet:zenodo} packages. \textsf{starry} models the stellar surface brightness and RV across the stellar disk by expanding them in spherical harmonics, a method that is well-suited for RM effect modelling, as described in \citet{Bedell2019}, \citet{Montet2020}, and \citet{Johnson2022}.

The joint photometric and RM-effect model enables the simultaneous fitting of all planetary parameters, ensuring that uncertainties are accurately propagated throughout the analysis.
The fitted parameters include the orbital period $P_f$, the planet-to-star radius ratio ($R_p/R_*$), the impact parameter $b$, the mid-transit time for photometry $T_0$ and for the RM event $T_{0,RM}$, and the quadratic limb-darkening coefficients $u_1$ and $u_2$. We included the fit of a quadratic trend to take into account any additional RV variations. The eccentricity and the mass of the planet were fixed to 0.059 and 25\Mearth, respectively, based on the RV analysis from HARPS-N, HIRES, HARPS, and ESPRESSO \citep[][Grouffal et al. in prep.]{Santerne2019}. The exact values of these parameters have minimal impact on the RM effect modelling. The priors used for each parameter are listed in Table \ref{tab_priors}.

To account for Transit Timing Variations, we fit separate mid-transit times for the photometry and RM datasets. RV offsets for each instrument measuring the RM effect were constrained by out-of-transit observations and overlapping coverage between instruments, ensuring well-constrained posteriors on the RV offsets.
For the RM effect fit, we place a broad uniform prior on the sky-projected obliquity $\lambda$, while stellar inclination is modelled following the method of \citet{2022ApJ...931L..15S}. The stellar radius $R_*$, the rotation period $P_{rot}$ and $\cos{i_*}$ are fitted simultaneously, with a uniform prior on $\cos{i_*}$ ranging from -1 to 1. The equatorial velocity is then derived as follows:

\begin{equation}
    v_{eq} = \frac{2\pi R_*}{P_{rot}}
\end{equation}

Using the methodology proposed by \citet{2020AJ....159...81M}, we estimate $ v \sin{i_*} = v_{eq}\sqrt{1 - \cos{i_*}^2}$. Finally, the 3D spin-orbit angle $\Psi$ is calculated using equation (9) of \citet{2009ApJ...696.1230F},

\begin{equation}
    \cos{\Psi} = \cos{i_*}\cos{i} + \sin{i_*}\sin{i}\cos{\lambda}.
\end{equation}

In \textsf{starry}, the adopted geometry of the system is such that the $x - y$ plane is the plane of the sky and the axis $z$ points toward the observer. The planet orbits around the star in a counterclockwise manner. In this context, the projected obliquity $\lambda$ is defined as the angle between the stellar rotation axis and the angular momentum vector of the planet projected on the sky plane \citep{Montet2020}. This convention differs from other RM models like \textsf{ARoME} \citep{Boue2013} and can induce opposite solutions for the value of $\lambda$. 

\begin{figure*}[h]
   \centering
   \sidecaption
   \includegraphics[width=0.7\textwidth]{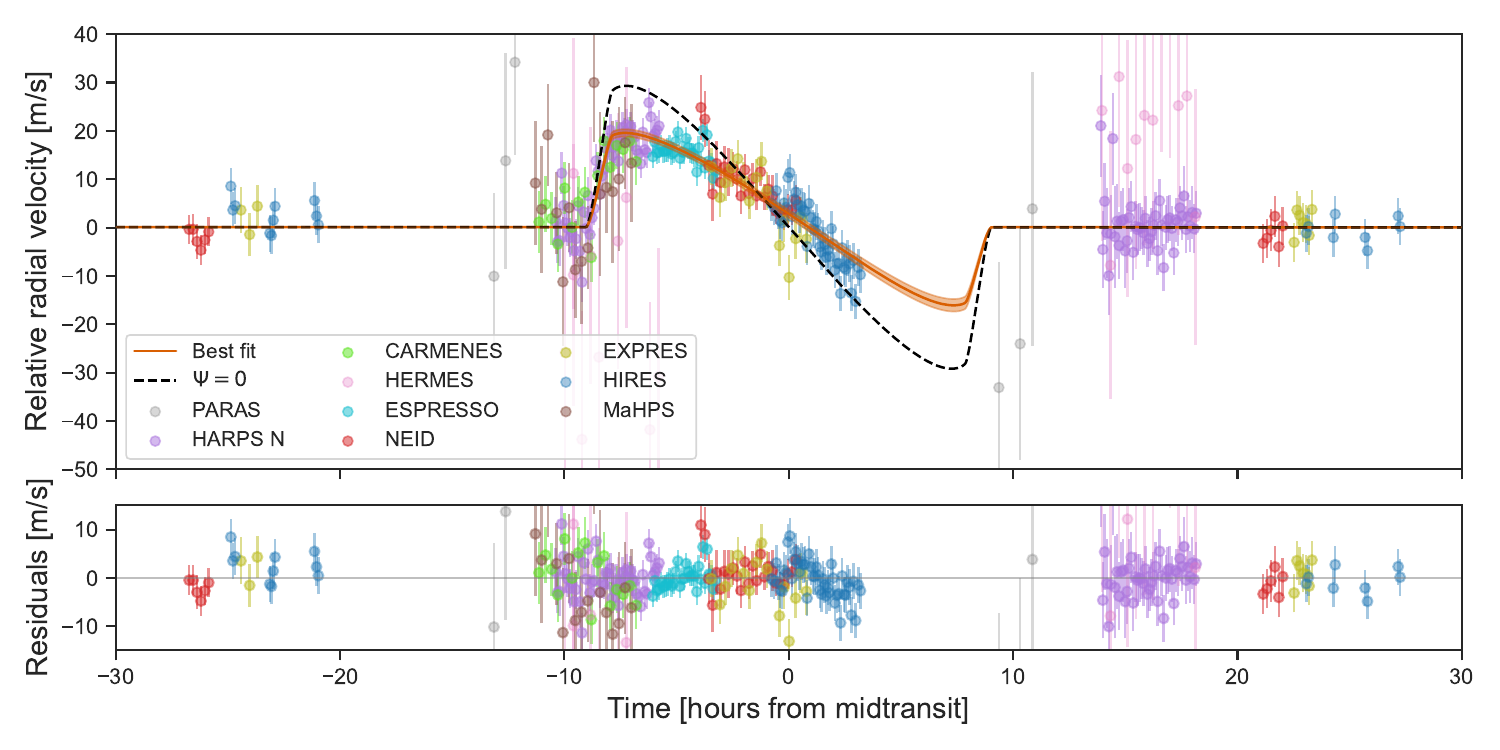}
      \caption{RM effect of the planet HIP~41378~f. Each instrument is represented by a colour: PARAS (grey), CARMENES (light green), HARPS-N (purple), HERMES (pink), ESPRESSO (light blue), NEID (red), HIRES (dark blue,) EXPRES (dark green) and MaHPS (brown). All instruments observed the same event. The median of the best fit is represented by an orange line, and the shadow area represents 1-$\sigma$ the uncertainties. The black dotted line corresponds to the RM effect of planet f if $\Psi =0$}
         \label{fig_final_fit}
   \end{figure*}

We ran a Markov Chain Monte Carlo (MCMC) simulation using the No-U-Turn Sampler implemented in PyMC3 \citep{Hoffman2011,Salvatier2015}. Two chains were run, each with 6000 tuning steps and 3000 posterior samples, with a target acceptance rate of 98 $\%$. All chains converged, and the Gelman-Rubin statistics for all parameters were less than 1.001, indicating reliable convergence \citep{Gelman1992}. 
Figure~\ref{fig_final_fit} presents the best-fit model together with the combined data from all instrument campaigns and the best fitted parameters are in Table \ref{tab_results}.

\begin{table}[h]
\caption{ Best-fitting main orbital parameters}\label{tab_parameters}%
\resizebox{\columnwidth}{!}{%
\begin{tabular}{@{}lcc@{}}
\hline
Parameter   & Values & Description       \\

\hline
$R_*$                  &  $1.299\pm 0.002$  &  Stellar radius ($R_{\odot}$)   \\
$R_p/R_*$               & $0.0666 \pm 0.0001$ & Radius ratio\\
$b$                     &  $0.18 \pm 0.01$ & Impact parameter\\
$T_0$                   &   $57186.9146\pm 0.0002 $ & Mid-transit time transit (BJD - 2400000)\\
$t_{0,RM}$              &  $59897.0199 \pm 0.0009$ & Mid-transit time RM (BJD - 2400000) \\
$P_f$                   & $542.0797_{-0.0002}^{+0.0001}$ & Orbital period (days) \\
$\lambda$               & $21 \pm 8$  & Sky-projected stellar obliquity (deg)\\
$a/R_*$                 &  $230.12_{-0.37}^{+0.35}$ & Scaled semi-major axis\\
$P_{rot}$                &$7.8 \pm 1.0$ & Stellar rotation period (days)\\
$i_*$                   &$42_{-7}^{+6}$ & Stellar inclination  (deg)\\
$v_{eq}\sin(i_*)$       &$5.6\pm 0.3$ & Projected rotational velocity  (\,km\,s$^{-1}$)\\
$v_{eq}$                & $8.4_{-0.8}^{+1.4}$ & Equatorial velocity  (\,km\,s$^{-1}$)\\
$\psi$                 & $52 \pm 6$ & True obliquity  (deg)\\
\hline
\end{tabular}%
}
\footnotetext{The full Table is available in Appendix}
\label{tab_results}
\end{table}

The inferred parameters, along with their uncertainties, are listed in Table \ref{tab_posteriors}. We find $T_{0,RM} = 2459897.0199 \pm 0.0009$ BJD. For the projected obliquity, we find $\lambda = 21\pm8$\degr\ with a stellar inclination of $i_* = 42_{-7}^{+6}$\degr\, leading to a 3D spin-orbit angle of $\Psi = 52 \pm 3$\degr. We found $v\sin i = 5.6 \pm 0.3$\kms, in agreement with the value found in \citet{Lund2019}. To determine whether the measured $v\sin i$ is incompatible with $\Psi = 0$\hbox{$^\circ$}, we analysed out-of-transit ESPRESSO spectra of the star (see figure \ref{fig_CCFanalysis}). Given the well-constrained values of the stellar rotation period and radius (Section \ref{subsec42}), a star with an inclination of 90 \hbox{$^\circ$} would have an equatorial velocity of $v_{eq} = 8.8 \pm 1.6 $ \kms. We compared the CCF of HIP~41378 with that of HD~8859, a star of similar type \citep{Sulis2023}, which has $v\sin(i_*)= 7.24 \pm 0.35$\,km\,s$^{-1}$. HD~8859 exhibits a broader CCF with a lower contrast. Additionally, we compared the data with a theoretical rotation profile corresponding to $v\sin(i_*)= 11$ \kms. Our analysis demonstrates that the ESPRESSO CCF of HIP~41378 is inconsistent with a higher $v\sin(i_*)$ required for $\Psi = 0$\hbox{$^\circ$}. While we did not account for instrumental broadening in the theoretical rotation profile due to its uncertainty, this omission does not affect our conclusion.

As the full RM effect is not observed, and that 30$\%$ of the transit is missing at egress time, the exact value of the projected obliquity is difficult to constrain and uncertainties in the derived value may be underestimated. Observations at egress time could help to confirm the value and reduce the uncertainties.

\subsection{Differential rotation}\label{subsec44}

In the previous analysis, we considered a solid rotation of the star. However, the rotation speed of a star can vary with latitude due to differential rotation. In systems where the projected obliquity $\lambda$ is non-zero, the planet transits across different stellar latitudes, and the RM signal would be sensitive to differential rotation.  

The stellar angular velocity at a given latitude $\theta$ is typically described by the equation:

\begin{equation}
    \omega = \omega_{eq} (1 - \alpha \sin^2{\theta}),
\end{equation}

where $\omega_{eq}$ is the equatorial angular velocity, and $\alpha$ is the rotational shear coefficient that quantifies the rotation rate between the equator and the pole. In the case of the Sun, $\alpha$ is equal to $0.2$. This formula is implemented in \textsf{starry}, allowing us to model the star's differential rotation. 

HIP~41378~f is an F-type star with $T_{eff} = 6290 \pm 77 K$ \citep{Lund2019}. \citet{2016MNRAS.461..497B} found that relative shear reaches a maximum for F-type stars and only $25 \%$ of F-type stars in their survey with no detectable shear. Shear is a complex function of both rotational rate and effective temperature, but according to the interpolation formula found by \citet{2016MNRAS.461..497B}, $\alpha$ should be between 0.08 and 0.2 for a star with $T_{eff} = 6290 \pm 77 K$ and a rotation period between 4.3 and 10.3 days. Moreover, \citet{Ammler-von2012} studied 180 stars of type A-F and found the largest relative differential rotation for F-type stars, encouraging us to explore the impact of differential rotation on HIP~41378. 

\citet{Gaudi2007} studied the detectability of differential rotation via the RM effect, concluding that for systems with stellar inclination $i_* = 60$\hbox{$^\circ$}, the optimal configuration for detection occurs when $\lambda \approx 70$\hbox{$^\circ$}. However, to achieve this, it requires a quality factor $Q_R = \sqrt{N} \frac{K_r}{\sigma} \gtrsim 200$, where $K_r$ is the RM amplitude, $N$ is the number of points and $\sigma$ is the measurement uncertainty. 
In most cases, this detectability threshold is difficult to achieve, especially during short-duration transits with limited data points. HIP~41378~f, with its 19-hour transit and excellent SNR, presents a unique opportunity. By using only the highest-quality instruments, we achieve a quality factor $Q_R \approx 190 $, close to the threshold for detecting differential rotation. This estimation is done without considering additional noise due to the combination of several instruments.

If HIP~41378 undergoes differential rotation, our calculation of the equatorial velocity $v_{eq}$ would be affected. It would depend on the latitude  $\theta$ of the stellar feature used to determine the stellar rotation period. For large values of the differential rotation parameter $\alpha$, the rotation period at the latitude of the stellar spot $P_{rot,spot}$ may differ substantially from the equatorial rotation period $P_{eq}$. This discrepancy can lead to an underestimation of $v_{eq}\sin i_*$, which in turn affects the projected obliquity fitted in the RM observations.

To account for potential differential rotation, we extended our model to include fits for both $\alpha$ and $\theta$. If the stellar feature used to deduce the rotation period is not located at the equator, then the relationship between $v_{eq}\sin i_*$ and $P_{rot,spot}$ becomes

\begin{equation}
    v_{eq}\sin i_* = \frac{2\pi R_* \sin i_*}{P_{rot,spot}(1-\alpha \sin^2\theta)}.
\end{equation}

In this extended model, we fit $P_{rot,spot}$, $R_*$ and $\cos i_*$ in the same way as in the global fit, adding a uniform prior on $\alpha$ (ranging from 0 to 1) and on $\theta$ (with $\cos(\theta)$ between 0 and 1). The equatorial velocity $v_{eq}$ is subsequently derived from $v_{eq}\sin i_*$ fitted on the RM-effect observations. With this approach, we obtain $\alpha = 0.93_{-0.10}^{+0.05}$ and $\theta = 20 \pm 9$\hbox{$^\circ$}, leading to $v_{eq}\sin i_* = 8.6_{-0.8}^{+0.9}$\kms, $\lambda = 38 \pm 3$ \degr, $i_* = 66 \pm 5$\hbox{$^\circ$} and finally $\psi = 45 \pm 3$\degr. The high value of $\alpha$ is not expected by studies of similar stars. Several effects could explain an overestimation of $\alpha$, like instrumental broadening or small shifts due to instrumental offsets. The differential rotation model better reproduces the slope of the HIRES data. According to \citet{Boue2013}, RVs derived using the CCF extraction method should show a steeper slope near mid-transit than RVs derived using the iodine cell, which is the inverse of what we observe for HIP~41378~f.

Finally, we decided to test a model based on studies of late F-type stars similar to HIP~41378. We performed a last analysis setting the differential rotation with $\alpha = 0.2$, similar to the solar value. It is not possible to have an independent constraint on the latitude of the spot and it is difficult to estimate the precise value of differential rotation for stars with $v\sin i < 20 $\,km\,s$^{-1}$ with the Fourier transform of spectral line profiles \citep{Takeda2020}. A non-zero value for the differential rotation that is not as high as the one found in the case where we set $\alpha$ free is therefore consistent with our knowledge of this type of star. We used the same method as described above with the same prior. Finally, we obtain $\theta = 38_{-20}^{+27}$\degr, leading to $v_{eq}\sin i_* = 6.3_{-0.5}^{+0.4}$\kms, $\lambda = 30_{-8}^{+7}$\degr, $i_* = 47 \pm 7$\degr and finally $\psi = 51 \pm 5$\degr. The values are all compatible with the fit without differential rotation.

The different models were compared using leave-one-out cross-validation (LOO-CV) as implemented in the ArviZ package. This method estimates the out-of-sample predictive performance of each model by systematically leaving out one data point at a time and evaluating how well the model predicts it. LOO-CV provides a robust, fully Bayesian approach to model comparison based on predictive accuracy.

We tested the three models with varying assumptions about stellar differential rotation. The model with free $\alpha$ yields the best predictive performance, with an expected log pointwise predictive density of –746.15, compared to –750.03 for $\alpha$ = 0.2 and –756.36 for $\alpha$ = 0. While this suggests a better fit, the improvement is not statistically significant given the large uncertainties ($\sim$17). Moreover, the differential rotation coefficient inferred in the free $\alpha$ model is large and seems inconsistent with realistic stellar rotation profiles of this type of star. We are likely at the sensitivity limit for constraining differential rotation with the current data, and a more detailed stellar model, or comparison with stars of similar type, is needed to robustly assess its presence. For these reasons, we present results assuming no differential rotation ($\alpha$ = 0) and present all detailed models in Table \ref{tab_posteriors}.

\subsection{Macroturbulence}
\label{sec_macroturbulence}

The code \texttt{starry} does not include the modelling of macroturbulence, which refers to the broadening of spectral lines due to convective motions in the outer layers of stars. For an F-type star such as HIP~41378, with an effective temperature of $T_\mathrm{eff} = 6290$ K \citep{Lund2019}, the macroturbulence can be non-negligible. \citet{Doyle2014} provide an empirical relation for the macroturbulent velocity $\zeta$, based on observations from \textit{Kepler}. Using the stellar parameters from \citet{Lund2019}, we estimate the macroturbulence velocity to be approximately $\zeta \approx 5.21$\kms. To assess the impact of both the CCF extraction method and macroturbulence, we performed an independent fit using only CCF-based instruments (HARPS-N, ESPRESSO, NEID, EXPRES) with the \texttt{ARoME} code \citep{Boue2013}. We set a uniform prior on $v \sin i_*$ (ranging from 2.5 to 20\kms) and on $\lambda$ (from $-180^\circ$ to $180^\circ$). The same prior as in the global analysis was used for the mid-transit time. Offset and jitter terms were fitted independently for each instrument. We adopted a Gaussian prior centred at 4.2\kms\ for the width of the intrinsic line profile of a non-rotating star. The macroturbulent velocity was assigned a normal prior centred at 5.2\kms\ with a standard deviation of 1\kms. The stellar rotation period was given a uniform prior between 6.0 and 8.5 days.

With \texttt{ARoME}, we obtain a projected obliquity of $\lambda_\mathrm{CCF} = 14 \pm 11$\degr\ and a projected rotational velocity of $v \sin i_*^\mathrm{CCF} = 6.4 \pm 1.0$\kms. The three-dimensional obliquity is $\psi_\mathrm{CCF} = 49^{+8}_{-11}$\degr. The fit converges toward a lower macroturbulence value, $\zeta = 3.5\pm1$ \kms, than the one predicted by the relation from \citet{Doyle2014}. Forcing the macroturbulence to that empirical value results in a poor fit to the data. The best fit is represented in Figure \ref{fig_ARoME}, and the parameters and priors are in Table \ref{tab_ARoME}.

All the results are compatible with the global fit with \textsf{starry} and \textsf{exoplanet}.

\subsection{Rossiter-McLaughlin Revolutions}

To complete the classical analysis of the RM effect, we conducted a complementary study using the RM Revolutions technique \citep[RMR,][]{Bourrier2021}, as implemented in the \textsc{antaress} workflow (\citealt{Bourrier2024}). This method makes direct use of the stellar lines occulted by the planet, exploiting their full spectral profile. Since it requires a reference spectrum of the unocculted star and stellar lines that do not vary temporally between the out-of-transit and in-transit exposures, we applied this technique exclusively on HARPS-N data, for which a pre-ingress baseline is available. 

\begin{figure}[h!]
   \centering
   \includegraphics[width=1.0\columnwidth]{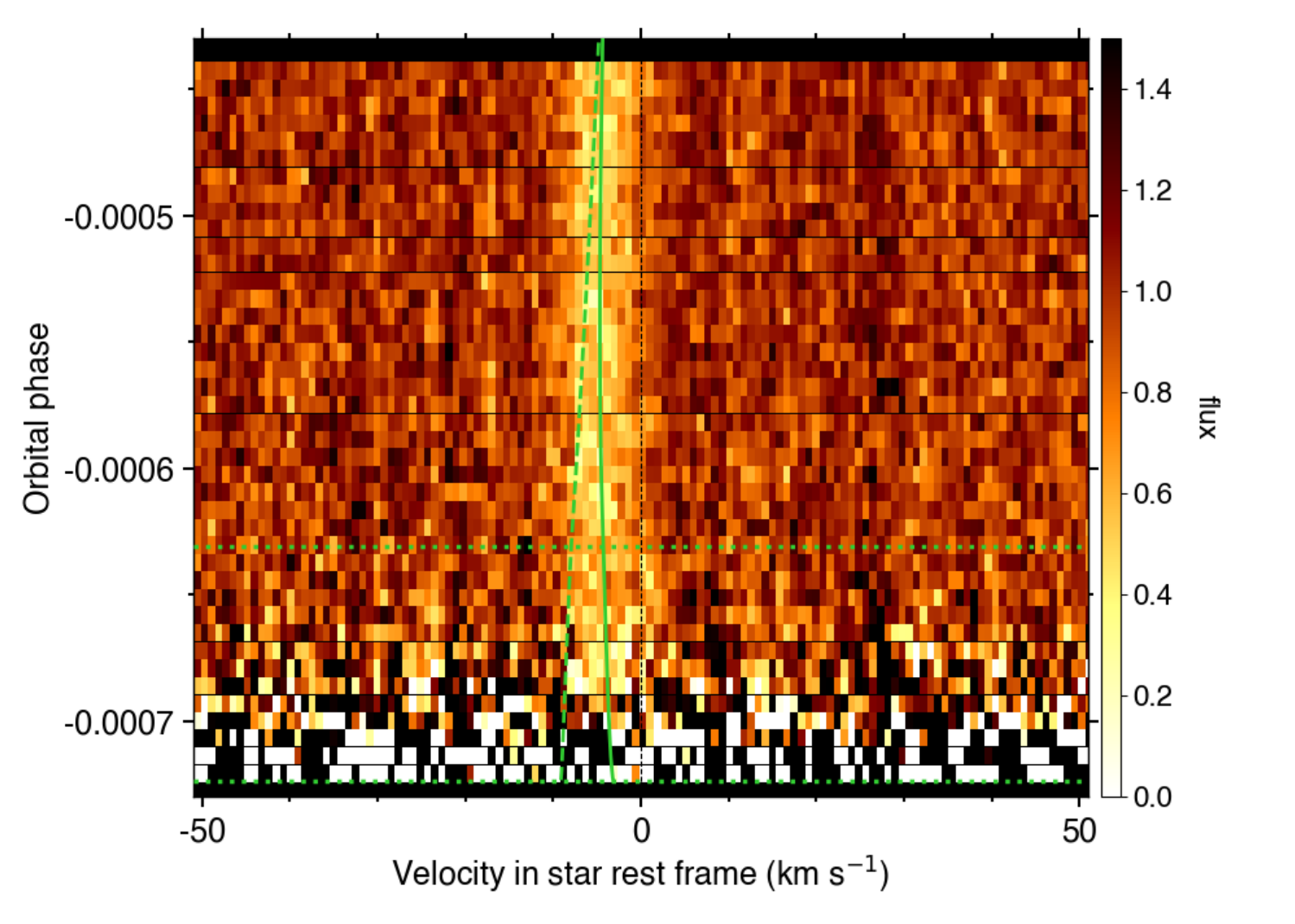}
      \caption{Map of the intrinsic CCF measured with HARPS-N during ingress. The solid green line represents the photospheric RV model from the RMR best fit, which deviates from a model with solid-body rotation (dashed green line) due to differential rotation. Ingress transit contacts are shown as green dotted lines.}
         \label{fig_RMR}
   \end{figure}

Disk-integrated CCFs were fitted with a Gaussian model to assess their quality. No systematic variations were found in the time-series of RVs, contrast, and full width at half maximum (FWHM), but their dispersion was reduced in the sky-corrected data, which was used thereafter. The weighted average of the out-of-transit RV residuals from the Keplerian model provides the systemic RV of the system for this dataset, and their combination was used to align all disk-integrated CCFs in the star rest frame. CCFs were then scaled to their correct relative flux level using a \textsc{batman} light curve (\citealt{Kreidberg2015}, generated using our derived transit depth and quadratic limb-darkening coefficients (Table \ref{posteriors}). This processing makes it possible to extract planet-occulted CCFs along the transit chord, which are then corrected for the planetary continuum absorption and local stellar intensity to obtain intrinsic CCFs tracing the pure spectral line profile of the stellar lines (Fig.~\ref{fig_RMR}).

\begin{figure}[h!]
   \centering
   \includegraphics[width=1.0\columnwidth]{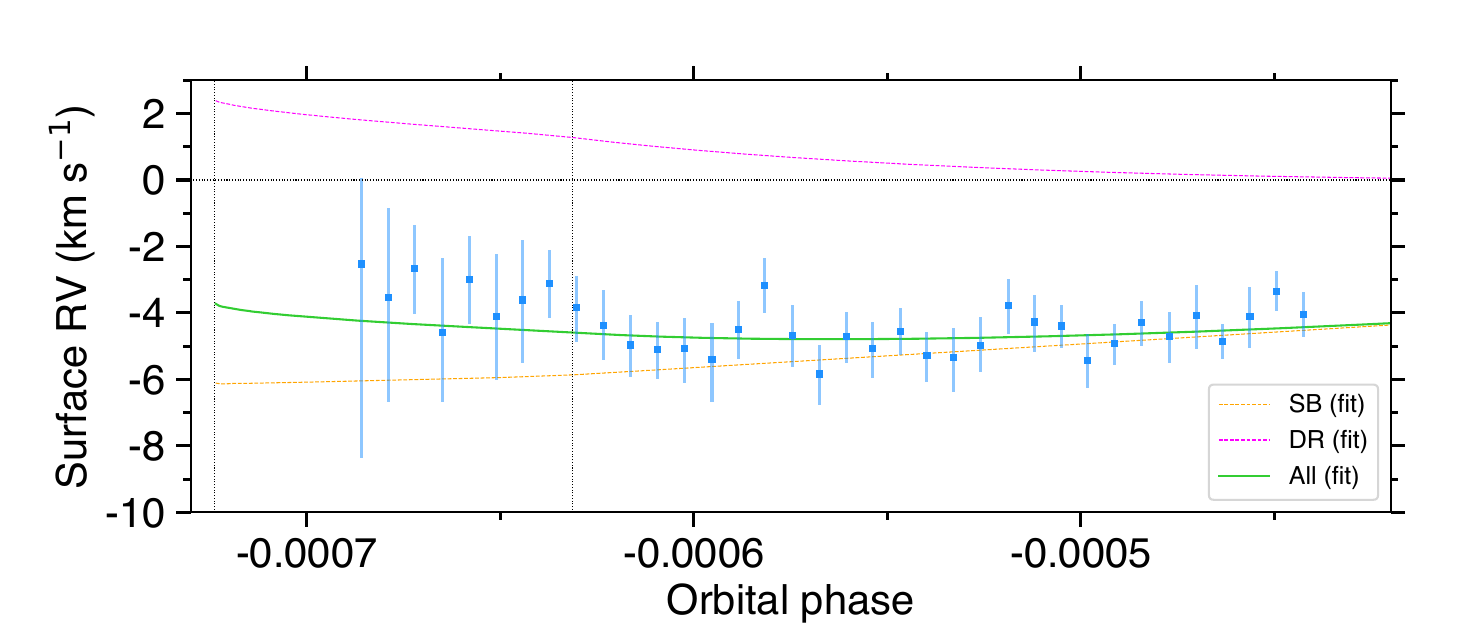}
      \caption{Photospheric RVs along the transit chord during ingress. Vertical dashed lines indicate ingress contacts. Blue points show the RV centroids derived from fits to individual CCFs. The green curve shows the best-fit model including differential rotation (DR), whose deviation with respect to solid-body (SB) rotation (orange curve) is plotted in magenta. }
         \label{fig_DR}
   \end{figure}

In a first step, intrinsic CCFs are fitted independently to assess their quality and to evaluate the best models describing the photospheric RV field, the stellar line profile, and its spatial variations across the photosphere. The intrinsic line of HIP 41378 is best modelled with a Gaussian profile. It is unconstrained in the first five in-transit exposures, measured at the faint stellar limb, which were subsequently excluded. Analyses of the intrinsic line properties favour a constant line contrast and FWHM along the probed transit chord. Surface RVs are better reproduced with a model including differential rotation but do not constrain convective blueshift (Fig.~\ref{fig_DR}).  

In a second step, we fitted a joint model of the stellar line, informed by the first step, to all intrinsic CCFs. The constraint on differential rotation breaks the degeneracy on the sky-projected stellar rotation. The joint RMR fit was thus performed using the stellar inclination (as its cosine $\cos i_\star$, with prior $\mathcal{U}$(-1, 1)) and equatorial rotational period $P_\mathrm{eq}$ as jump parameters, together with the relative differential rotation rate $\alpha$ ($\mathcal{U}$(0,1)), sky-projected spin-orbit angle $\lambda$ ($\mathcal{U}$(-180,180)$^{\circ}$), and intrinsic line contrast ($\mathcal{U}$(0,1)) and FWHM ($\mathcal{U}$(0,10)\,km\,s$^{-1}$, with upper limit set by the disk-integrated line FWHM). A normal prior informed by the stellar rotation analysis (Sect.~\ref{subsec42}) was set on $P_\mathrm{eq}$, providing consistent and refined results compared to using a broad uniform prior. 

The HARPS-N data allow for two possible configurations: one with the stellar north pole visible ($i_* < 90$\hbox{$^\circ$}) and positive $\lambda$, and the other with the south pole visible ($i_* > 90$\hbox{$^\circ$}) and positive $\lambda$. In both cases, the best-fit favours high values for $\alpha$, reinforcing the findings from the classical analysis. While the HARPS-N data alone favour the southern configuration, we isolate the northern one for consistency with the classical analysis of all datasets. It corresponds to $i_* = 73_{-7}^{+6}$\hbox{$^\circ$}, $\lambda = 35.0 \pm 3.0$\hbox{$^\circ$}, and $\alpha > 0.85$ at the 1$\sigma$ level. Combining the PDF for $\lambda$, $i_*$, and the orbital inclination yields a 3D spin-orbit angle $\Psi = 38.7 \pm 4.2$\hbox{$^\circ$}.

Despite a limited instrumental coverage, the RMR analysis of the HARPS-N dataset independently confirms the classical results and reinforces the conclusion that the system is likely to be misaligned. The results are compatible with the classical analysis fit with $0<\alpha<1$ at less than 1.6-$\sigma$. It supports the presence of stellar differential rotation, although the magnitude of this effect remains challenging to explain and suggests that more refined differential rotation models may be required for accurate RM-effect modelling. A more precise study is needed, particularly through comparison with similar stars, which we leave for future work. 

\section{Discussion}
\label{discussion}

\begin{figure}[!h]
   \centering
   \includegraphics[width=0.7\columnwidth]{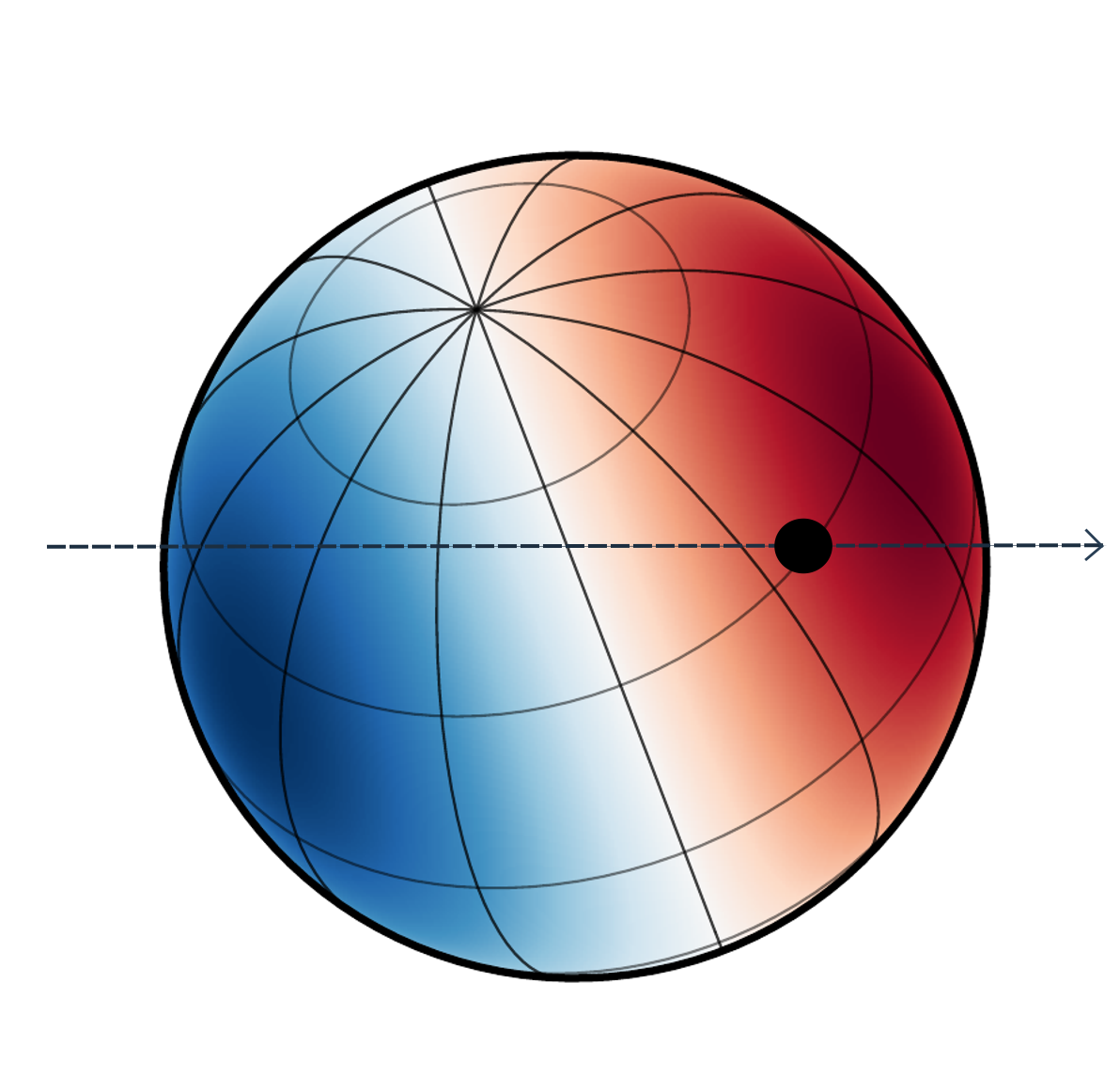}
      \caption{ View of the system with \textsf{starry} package without differential rotation. The obliquity and stellar inclination have been set to the median of the posterior from the RM analysis. The colour represents the blueshift and redshift of the star during its rotation on itself. The size of the planet and its position in the horizontal direction is arbitrary. }
         \label{fig_visualisation_system}
   \end{figure}

HIP~41378~f stands out as the longest-period planet for which the RM effect has been observed so far. Using nine instruments across four continents, we detected more than half of the transit through spectroscopy, including a clear signal of the ingress and the RM effect slope. This high SNR led to precise measurement of the transit time and provided detailed insights into the system's architecture. HIP~41378~f has a low impact parameter but is incompatible with zero, allowing for constraints on the parameters $\lambda$ and $v sini_*$ without introducing a strong correlation between them, as suggested by \citet{Gaudi2007}. A graphical representation of the system configuration is provided in Figure \ref{fig_visualisation_system}.

\subsection{A likely misaligned planetary system}

Previous RM studies focused mainly on short-period planets ($<$ 100 days), which exhibit a wide range of obliquities (see figure \ref{fig_stats_obliquities}). However, HIP~41378~f belongs to a sparse but growing population of long-period, slightly misaligned planets, together with HD~80606~b \citep{Moutou2009, Hébrard2010} and  TIC 241249530b \citep{Gupta2024}, with periods of 111 and 165 days, and eccentricities of 0.93 and 0.94 (respectively). TIC 241249530b is considered a potential progenitor of hot Jupiters \citep{Debras2021}. HIP~41378~f, with its low eccentricity and multi-planet system \citep{Santerne2019}, certainly followed a different evolution history. This observation, therefore, enables us to test other theories of planetary formation and evolution.

Stars hotter than the Kraft break \citep{1967ApJ...150..551K} are more frequently associated with misaligned systems, potentially due to differences in their internal structures \citep{2010ApJ...718L.145W, 2024ApJ...968L...2L}. With $T_{eff} = 6290$ K, HIP~41378 confirms this trend. The Kraft break dichotomy is generally associated with a difference in tidal dissipation affecting hot Jupiters and planets close to their stars. Observing long-period planets that are not affected by tides is crucial for determining whether tidal dissipation is responsible for the difference in misalignment between cool and hot stars. 

The obliquity of HIP 41378 measured from the transit of planet f is fully compatible with that of planet d ($\Psi_d = 69_{-11}^{+15}$\hbox{$^\circ$}), confirming that this vast transiting multi-planetary system has likely near-coplanar orbits \citep{Grouffal2022}. This similarity in obliquity is expected since all the planets are transiting. Therefore, the entire system is very likely misaligned with the host star. However, we note that the obliquity measurement for planet d was obtained from a RM observation with low SNR and may not be fully reliable \citep{Grouffal2022}. This uncertainty should be taken into account when interpreting the consistency between the two measurements. Future observations of the RM effect for planet e, which has a larger radius, once its orbital period is firmly established, could provide an additional, more robust comparison point to confirm the overall misalignment of the system.

The presence of a massive, inclined outer planet can tilt inner planets with nodal precession operating on a million-year timescale. This mechanism has been proposed to explain the misalignment of HAT-P-11 b, for example \citep{Yee2018}. However, with seven years of RV monitoring \citep[][Grouffal et al. in prep.]{Santerne2019}, we detected no signs of a very long-period planet more massive than 0.15 $M_{Jup}$ at a period longer than 2000 days. We, therefore, excluded perturbation from another external planetary companion to explain the tilt of HIP~41378 d and f.

As postformation misalignment theories seem disfavoured for this system, the host star HIP~41378 most likely has a primordial misalignment with its protoplanetary disk from which the planets formed. Based on \citet{2022PASP..134h2001A}, there are three main scenarios to explain a primordial misalignment, which are discussed below.

Magnetic interactions between the young star and its protoplanetary disk may explain the observed misalignment, where differential rotation between the star and the inner disk creates a tilt, amplified by the Lorentz force \citep{Foucart2013}. Observations of misaligned pre-main-sequence stars with protoplanetary disks \citep{Davies19} and obliquity measurements of debris disks \citep{Hurt23} also support this hypothesis. This scenario can reproduce a large diversity of obliquity, as seen from the observations \citep{Knudstrup2024}. This is also compatible with hot Jupiters' misalignment, as post-formation tidal damping could affect the obliquity of these planets. In this case, this would imply either that the disk is rigid from the inner region that experiences the shear to the orbital period of HIP~41378~f, or that different disk components are misaligned relative to each other, as shown in \citep{Francis2020}, but at different scales.
Another potential scenario is chaotic accretion, where interaction with clumps of gas or stellar companions in a turbulent star-forming region can cause the material to infall obliquely \citep{Fielding2015, Bate2018}. However, chaotic accretion is unlikely for HIP~41378~f as it tends to keep stellar obliquity within 20 degrees \citep{Takaishi20}, which is not the case for HIP~41378~f.
The presence of a tilted outer companion during the disk formation phase could have induced a misalignment of the protoplanetary disk or the stellar spin axis. A search in the Gaia Archive reveals no resolved binary companions within approximately 1 arcsecond of HIP 41378. While an unresolved companion could in principle be responsible for the observed tilt, the Renormalised Unit Weight Error (RUWE) for HIP 41378 is 0.982, a value very close to 1, indicating that the star is likely single and that the astrometric solution is reliable. A difference in amplitude of the RM effect or transit depth according to the wavelength could indicate the presence of a companion star. We do not observe differences between HARPPS-N and CARMENES, which were observed at the same time but with different wavelength ranges. Similarly, the transit of planet f observed with HST \citep{Alam2022} is compatible with the transit observed by K2 \citep{Vanderburg2016}. No sign of a stellar companion has been detected, but this hypothesis cannot be discarded without more observations.

Among the three scenarios, magnetic warping would be a viable explanation for the likely misalignment of HIP~41378's planetary system. Expanding the sample of long-period planets with obliquity measurements would help to test the Kraft break hypothesis and see if colder stars like the Sun also show misaligned systems.

HIP~41378 is not the only multi-planetary system where a primordial origin is suspected for spin–orbit misalignment. A notable example is K2-290 A, a star in a hierarchical triple stellar system, which hosts two coplanar planets with orbital periods of 9.2 and 48.4 days. Despite their coplanarity, the system exhibits a significant spin–orbit misalignment of $124\degr \pm 6\degr$ between the stellar equator and the planetary orbits \citep{Hjorth2021}. In this case, the strong dynamical influence of the two stellar companions is expected to be the cause of this misalignment \citep{Hjorth2021, Best2022}.

\subsection{Resonances and spin-orbit angles}

\begin{figure*}[h!]
	\centering
	\includegraphics[width=1.0\textwidth]{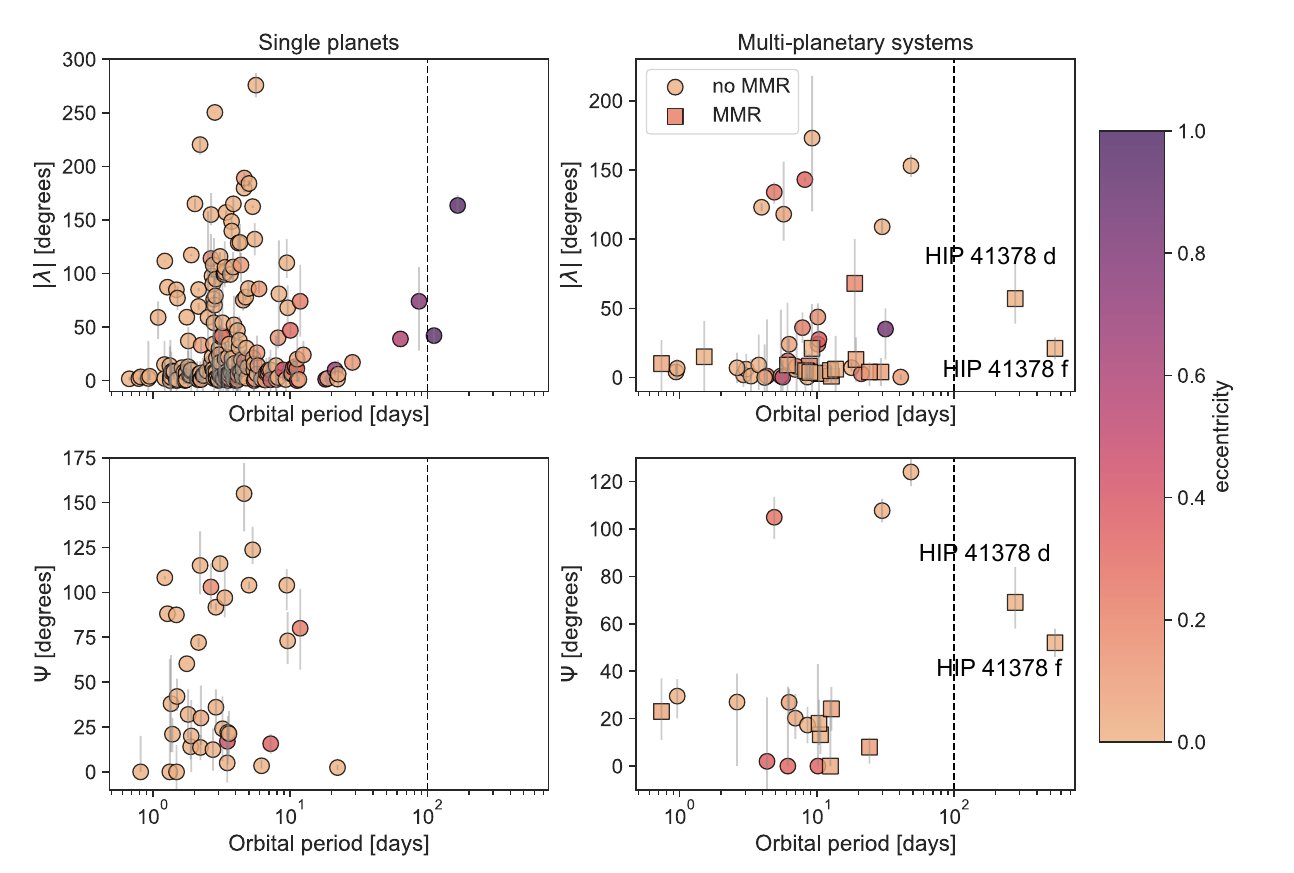}
	   \caption[Obliquities in confirmed exoplanets]{Exoplanets with measured projected obliquity $\lambda$ (upper panels) and true obliquity $\psi$ (lower panels) as a function of their orbital period. Left panels show systems with a single detected planet, while right panels show planets in multi-planet systems. The color map indicates orbital eccentricity, with darker shades representing higher eccentricities. For planets in multi-planet systems, squares indicate those near a mean motion resonance (MMR). A system is considered near MMR if the orbital period ratios of adjacent planets are close to small integer ratios (e.g., 2:1, 3:2), based on values reported in the literature for each system. These criteria are derived from the period data and resonance assessments in the respective discovery or follow-up studies. Data are taken from the TEPCat catalogue \citep{Southworth2011}, accessed on 26 February 2025.}
		  \label{fig_stats_obliquities}
	\end{figure*}

Beyond the measured spin-orbit misalignment, any theory aiming to explain the HIP~41378 system's architecture must also account for its multiplicity and resonant structure. The two inner planets, HIP~41378 b and c, orbit at 15.6 and 31.7 days, respectively, placing them near a 2:1 mean motion resonance (MMR) \citep{Vanderburg2016}. The outer planets lie in or close to a 4:3-3:2 resonant chain \citep{Santerne2019}, suggesting an ordered orbital configuration. A system in resonant configuration has also been proposed to explain the unusually low density of HIP~41378~f, potentially linked to the presence of circumplanetary rings \citep{Lu2025}. Resonant configurations are generally associated with smooth formation processes, making it unlikely that violent dynamical interactions contributed to the observed misalignment \citep{Rice23}. This strengthens the case for a primordial origin of the system's tilt, as dynamical processes capable of producing large obliquities typically disrupt resonances \citep{ Izidoro17, Esteves23}. 
HIP~41378 is not the only multi-planet system known to exhibit misalignment. In the Kepler-56 system, two close-in, coplanar planets with orbital periods of 10.50 and 21.41 days were found to be misaligned by approximately 45\hbox{$^\circ$}, based on a combination of asteroseismology and dynamical modelling \citep{Huber2013}. The detection of a distant outer companion in that system supports the hypothesis that perturbations from an inclined outer planet may have driven the inner planets' tilt. A similar scenario has been proposed for Kepler-129, a system composed of two planets at 15.79 and 82.20 days. There, a stellar inclination of $\sim$38\hbox{$^\circ$} has been inferred, suggesting a misalignment that may also result from the influence of a massive companion. Another case is HD~3167, a system in which the innermost planet is nearly aligned with the stellar rotation axis, while the two outer planets (c and d) are found to orbit in a plane almost perpendicular to that of the inner planet \citep{Dalal2019, Bourrier2021}. As discussed in \citet{Grouffal2022}, this configuration has a very low probability for the HIP~41378's planetary system considering the low mutual inclination between the planets and the long periods of the planets. Figure \ref{fig_stats_obliquities} summarises the current sample of confirmed exoplanets with measured obliquities. The low projected obliquity of HIP~41378~f is consistent with the findings of \citet{Albrecht2013}, which suggest that planets in multi-planet systems tend to be well-aligned. However, the measured 3D obliquity of HIP~41378~f deviates from previously observed values in such systems. To date, only a handful of multi-planet systems have obliquity measurements, and these are only for planets with orbital periods shorter than 50 days. Additional observations of long-period planets, particularly using a combination of instruments, would help to better interpret the measurement for HIP~41378~f and provide further constraints on the formation and evolutionary pathways of long-period exoplanets.  In particular, comparing more systems with planets near or far from mean-motion resonances is needed to determine whether different evolutionary pathways are at play.

\subsection{Combination of different instruments and models' limitations}

Studying the RM effect of HIP~41378~f requires combining data from multiple instruments, as using only a few instruments would significantly reduce the ability to derive a precise transit time and accurately measure the projected obliquity and $v \sin{i_*}$ values. However, each instrument and its RV extraction process can influence the RM effect, potentially impacting results. 

\citet{Boue2013} showed that instruments using a Gaussian fit of the CCF or the iodine cell technique can yield different RM-effect signatures. To verify consistency, we compared our \textsf{starry}-based fit with one produced by the \textsf{ARoME} code \citep{Boue2013}, which is designed for CCF-based instruments, in section \ref{sec_macroturbulence}. For this, we modelled HARPS-N, ESPRESSO, NEID, and EXPRES data together, as they share a common RV extraction method and high precision. We also included macroturbulence in the model.

Despite the challenges of combining datasets, all analyses, even including macroturbulence, consistently indicate that the projected obliquity of HIP~41378~f is low but different from 0 and that the true obliquity is significant. That result could indicate a misaligned system. \citet{Boue2013} found that RM slopes are typically steeper for CCF than for iodine cell data, whereas we observe the opposite for HIRES. If this discrepancy is due to instrumental effects, our only option is to analyse HIRES separately, as done in the \textsf{ARoME} analysis.

In addition to instrumental considerations, stellar phenomena can also impact the RM waveform. Our attempt to model differential rotation revealed large values of the rotational shear coefficient, suggesting that more complex stellar effects may influence the measurement. In particular, centre-to-limb variations and convective blueshift of spectral lines, as described by \citet{Cegla2016}, can significantly distort the RM signal, especially in rapidly rotating stars like HIP~41378 ($v \sin{i_*} = 5.6 \pm 0.3$ \kms). Neglecting these effects can introduce biases of up to $20\degr$ in the derived obliquity, potentially reducing the inferred true obliquity of HIP~41378~f to $\sim30\degr$, which still supports a misaligned configuration. Our current \textsf{starry} model does not include instrumental line broadening, nor does it account for these convective and limb-darkening complexities. Future analyses of high signal-to-noise RM datasets, such as this one, will benefit from the inclusion of such effects to improve the accuracy of obliquity measurements. 
These considerations may help to explain some of the structures observed in the residuals. In particular, although a fit without HIRES data is compatible with the entire analysis in terms of the system's parameters, the difference in RV extraction methods could explain the residuals of the HIRES data.

\section{Conclusions}
\label{conclusion}

The RM effect of HIP~41378~f was significantly detected during its transit on 12–13 November 2022. Our analysis indicates that the planet follows a prograde orbit, potentially misaligned with the stellar spin axis.
Our results show that HIP~41378~f has a slightly misaligned projected obliquity of $21\pm8$\degr. Although all analyses converge toward a projected obliquity between 10 and 30 degrees, the egress part of the transit is missing. It is therefore possible that the derived uncertainties are overly optimistic. However, even in the case of a projected obliquity close to zero, the values of $v_\mathrm{eq} \sin i_*$ obtained from two independent analyses using \texttt{starry}, including differential rotation, and \texttt{ARoME}, including macroturbulence, are inconsistent with a stellar inclination of $i_* = 90^\circ$. Both the photometric and RV analyses converge toward a stellar rotation period between 6 and 9 days. A longer rotation period of 11 days, which would imply a nearly aligned system, is not supported by any dataset. The three-dimensional obliquity of the system is non-zero. However, we acknowledge that this conclusion relies on the measured stellar rotation period. Combining the observed RM signal with precise stellar parameters, we determined a 3D spin-orbit angle of $\Psi = 52 \pm 6$\hbox{$^\circ$}.

We also investigated the influence of differential rotation, a common feature in late F-type stars such as HIP~41378. Our results suggest that the RM signal is compatible with stellar differential rotation. However, we emphasise that other stellar effects, such as convective blueshift, and centre-to-limb variations can also alter the RM waveform, especially at high signal-to-noise, and should be accounted for in future analyses.

While HIP~41378 is not the first misaligned multi-planetary system, it is unique due to the long orbital period of planet f. It now represents the longest-period exoplanet for which the obliquity has been measured via the RM effect. Systems like HIP~41378, with transiting long-period planets, provide exceptional opportunities to probe the dynamical histories of planetary systems and test models of planet formation and migration over wide spatial scales. Considering the number of planets in the system and resonances between them, a primordial misalignment of the star is favoured, maybe involving early magnetic wrapping of the star.

Finally, this study highlights the feasibility of combining observations from multiple spectrographs to analyse long-duration transits. By coordinating instruments across different observatories, we achieved precise parameter determinations from RM observations, showing that this approach could benefit future studies of long-period exoplanets.

Looking ahead, the upcoming PLATO mission \citep{Rauer2014}, which aims to uncover and characterise long-period planets, will offer a unique opportunity to validate and refine theories of planetary formation in diverse environments. International collaboration, like for HIP~41378 f, will help to follow-up systems that PLATO will discover.

\begin{acknowledgements}
We thank the anonymous referee for providing insightful comments on the manuscript, which enabled us to improve the final version. The project leading to this publication has received funding from the Excellence Initiative of Aix-Marseille University - A*Midex, a French “Investissements d’Avenir programme” AMX-19-IET-013. This work was supported by the "Programme National de Planétologie" (PNP) of CNRS/INSU co-funded by CNES. O.B-R. and J.L.-B. are funded by the Spanish Ministry of Science and Universities (MICIU/AEI/10.13039/501100011033) and NextGenerationEU/PRTR grants PID2019-107061GB-C61, PID2023-150468NB-I00 and CNS2023-144309. These results made use of the Lowell Discovery Telescope at Lowell Observatory. Lowell is a private, nonprofit institution dedicated to astrophysical research and public appreciation of astronomy and operates the LDT in partnership with Boston University, the University of Maryland, the University of Toledo, Northern Arizona University, and Yale University. The EXPRES team acknowledges support for the design and construction of EXPRES from NSF MRI-1429365, NSF ATI-1509436 and Yale University. D.A.F. gratefully acknowledges support to carry out this research from NSF 2009528, NSF 1616086, NSF AST-2009528, the Heising-Simons Foundation, and an anonymous donor in the Yale alumni community. We acknowledge financial support from the Agencia Estatal de Investigaci\'on of the Ministerio de Ciencia e Innovaci\'on MCIN/AEI/10.13039/501100011033 and the ERDF “A way of making Europe” through project PID2021-125627OB-C32, and from the Centre of Excellence “Severo Ochoa” award to the Instituto de Astrofisica de Canarias. NCS acknowledges funds by the European Union (ERC, FIERCE, 101052347). Views and opinions expressed are however those of the author(s) only and do not necessarily reflect those of the European Union or the European Research Council. Neither the European Union nor the granting authority can be held responsible for them. This work was supported by FCT - Fundação para a Ciência e a Tecnologia through national funds by these grants: UIDB/04434/2020, UIDP/04434/2020. This research was funded in whole or in part by the UKRI, (Grants ST/X001121/1, EP/X027562/1). X.D acknowledges the support from the European Research Council (ERC) under the European Union’s Horizon 2020 research and innovation programme (grant agreement SCORE No 851555) and from the Swiss National Science Foundation under the grant SPECTRE (No 200021$\_$215200). This work has been carried out within the framework of the NCCR PlanetS supported by the Swiss National Science Foundation under grants 51NF40$\_$182901 and 51NF40$\_$205606. SCCB acknowledges the support from Fundação para a Ciência e Tecnologia (FCT) in the form of work contract through the Scientific Employment Incentive program with reference 2023.06687.CEECIND. PJW acknowledges support from the UK Science and Technology Facilities Council (STFC) through consolidated grants ST/T000406/1 and ST/X001121/1. We acknowledge the generous support from PRL-DOS (Department of Space, Government of India) and the Director, PRL for the PARAS-1 spectrograph funding for the exoplanet discovery project and research grant for SB. We thank Prof. Abhijit Chakraborty (PI, PARAS-1 spectrograph) for providing instrument observing time for this work. We also acknowledge the help from the PARAS-1 instrument teams and the Mount-Abu observatory staff for their assistance during the observations. A.M. acknowledges funding from a UKRI Future Leader Fellowship, grant number MR/X033244/1 and a UK Science and Technology Facilities Council (STFC) small grant ST/Y002334/1. S.G.S acknowledges the support from FCT through Investigador FCT contract nr. CEECIND/00826/2018 and  POPH/FSE (EC). This work has been carried out within the framework of the NCCR PlanetS supported by the Swiss National Science Foundation under grants 51NF40$\_$182901 and 51NF40$\_$205606. This project has received funding from the European Research Council (ERC) under the European Union's Horizon 2020 research and innovation programme (project {\sc Spice Dune}, grant agreement No 947634). BA acknowledges support of the Swiss National Science Foundation under grant number PCEFP2\_194576. This work was supported by Fundação para a Ciência e a Tecnologia (FCT) through the research grants [UID/FIS/04434/2019,] UIDB/04434/202, UIDP/04434/2020 and  2022.06962.PTDC (http://doi.org/10.54499/2022.06962.PTDC). The Wendelstein 2.1 m telescope project was funded by the Bavarian government and by the German Federal government through a common funding process. Part of the 2.1 m instrumentation including some of the upgrades for the infrastructure were funded by the Cluster of Excellence “Origin of the Universe” of the German Science foundation DFG.LT acknowledges support from the Excellence Cluster ORIGINS funded by the Deutsche Forschungsgemeinschaft (DFG, German Research Foundation) under Germany's Excellence Strategy – EXC 2094 – 390783311. SG is supported by an NSF Astronomy and Astrophysics Postdoctoral Fellowship under award AST-2303922. C.K.H.\ acknowledges support from the National Science Foundation (NSF) Graduate Research Fellowship Program (GRFP) under Grant No.~DGE 2146752.
\end{acknowledgements}

\bibliographystyle{aa} 
\bibliography{aanda}

\begin{appendix}
\onecolumn
\section{Stellar rotation period analysis}\label{secA1}
\FloatBarrier

      \begin{figure*}[h!]
   \centering
   \includegraphics[width=1.\textwidth]{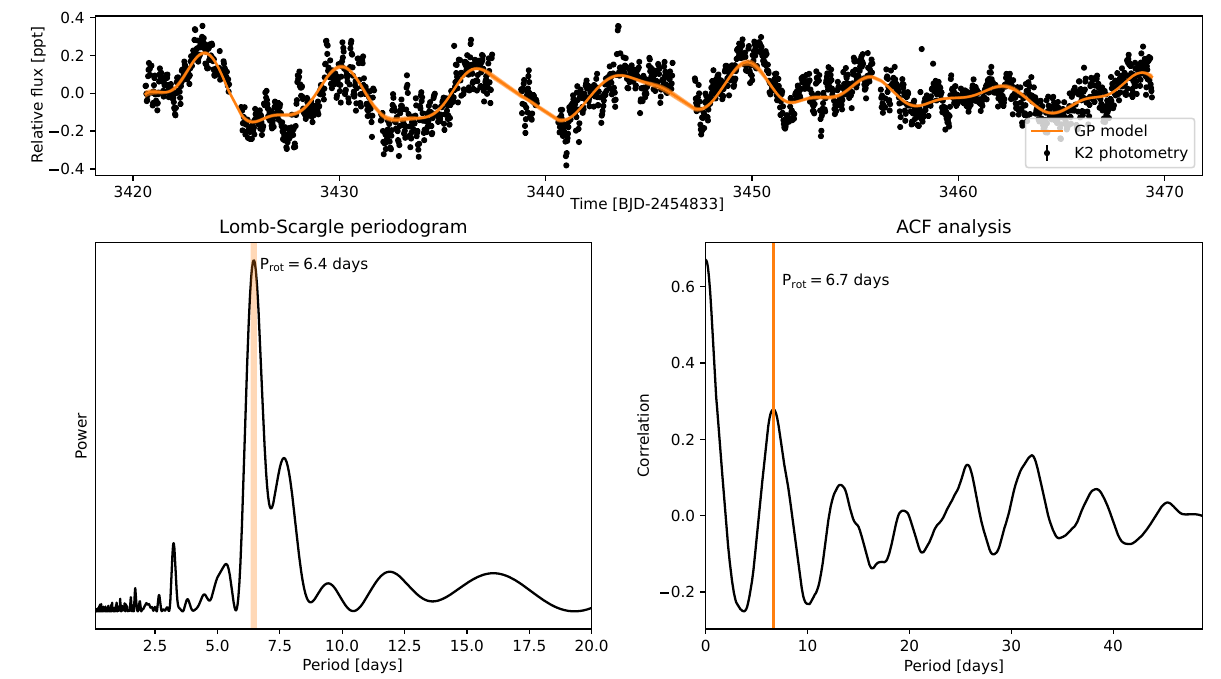}
      \caption{\textit{Upper panel:} K2 C18 data with all the transits removed. The best-fit of the rotation period with a GP model is represented as an orange line.\textit{Lower left panel:} Generalised Lomb-Scargle Periodogram. \textit{Lower left panel:} Autocorrelation analysis. The maximum power is highlighted in orange. }
         \label{fig_rotation_period}
   \end{figure*}

\section{ESPRESSO out-of-transit CCF}\label{secA2}
\FloatBarrier

      \begin{figure*}[h!]
   \centering
   \includegraphics[width=0.6\columnwidth]{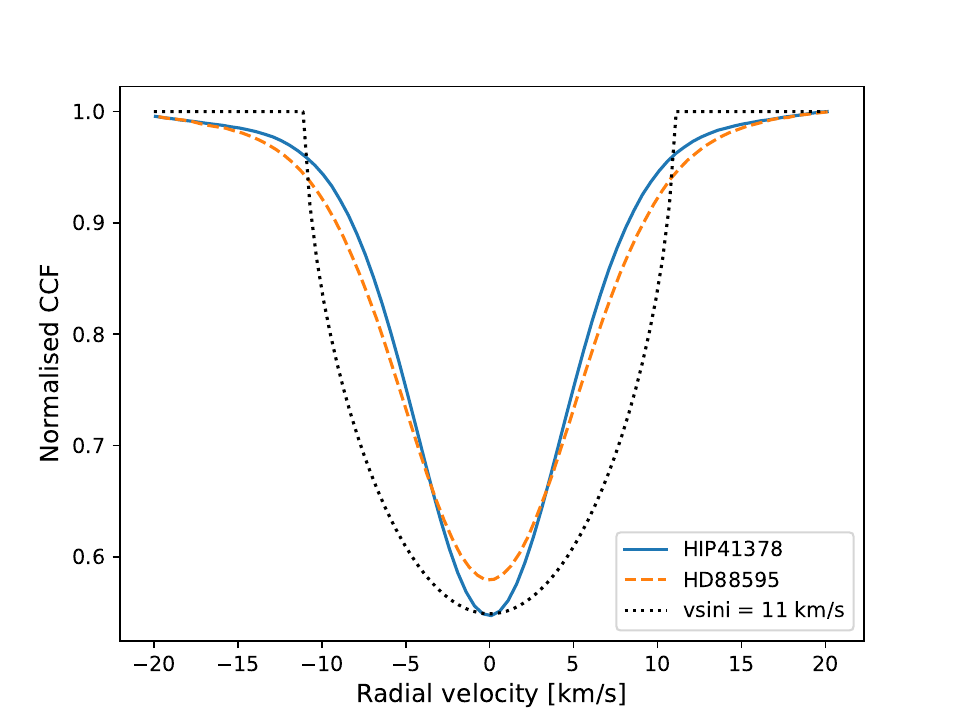}
      \caption{Out-of-transit ESPRESSO CCF of HIP~41378 in blue, compared with one CCF of HD~88595 \citep{Sulis2023}, a star similar in $T_{eff}$ to HIP~41378 with a higher $v\sin(i_*)= 7.24 \pm 0.35$ \,km\,s$^{-1}$ as an orange dashed line. The theoretical rotation profile of a star with $v\sin(i_*)= 11 $ \,km\,s$^{-1}$ is represented as a black dotted line. }
         \label{fig_CCFanalysis}
   \end{figure*}

\FloatBarrier
\section{Summary of the priors and posteriors for the RM-transit Analysis}

\begin{table}[hbt!]
\setlength{\tabcolsep}{50pt}
\caption{ List of parameters and priors used in the classical analysis.}\label{tab_priors}
\begin{tabularx}{\textwidth}{@{} llc @{}}
\hline
Parameter   & Description                      & Prior           \\
\hline
\textit{MCMC Inputs:} &       &               \\
$R_*$      & Stellar radius ($R_{\odot}$) &   $\mathcal{N}(1.300, 0.009)$                          \\
$u_1$      & Quadratic Limb-darkening coeff & \citet{Kipping2013}       \\
$u_2$      & Quadratic Limb-darkening coeff & \citet{Kipping2013}          \\
$R_p/R_*$  & Radius ratio               &  $\mathcal{U}(0.0, 0.2)$       \\
$b$       & Impact parameter            &   $\mathcal{U}(0, 1+R_p/R_* )$                          \\
$T_0$    & mid-transit time transit (BJD - 2400000) & $\mathcal{N}(57186.91, 0.01)$\\
$T_{0,RM}$    & mid-transit time RM (BJD - 2400000)    & $\mathcal{N}(59897.04, 0.01)$  \\
$P_f$     & Orbital period (days)       &  $\mathcal{N}(542.0797, 1)$     \\
$P_{rot}$     & Stellar rotational period (days)    &  $\mathcal{U}(6.0, 9.0)$    \\
$\lambda$     & Sky-projected stellar obliquity (rad)       &  $\mathcal{U}(-\pi,+\pi )$     \\
$\cos{(i_*)}$     & cosinus stellar inclination       &  $\mathcal{U}(0,1 )$     \\
$\alpha$     & rotational shear coefficient       &  $\mathcal{U}(0,1 )$     \\
$cos(\theta)$ & cosinus latitude of the stellar spot & $\mathcal{U}(0,1 )$ \\
$\gamma_{C18}$     & K2 C18 offset (ppm)     &  $\mathcal{N}(1,10 )$   \\
$\gamma_{HARPS-N}$     & HARPS-N offset (\,m\,s$^{-1}$)     &  $\mathcal{U}(-60, +60 )$   \\
$\gamma_{ESPRESSO}$   & ESPRESSO offset (\,m\,s$^{-1}$)     &  $\mathcal{U}(-60, +60 )$   \\
$\gamma_{CARMENES}$   & CARMENES offset (\,m\,s$^{-1}$)     &  $\mathcal{U}(-60, +60 )$  \\
$\gamma_{NEID}$   & NEID offset (\,m\,s$^{-1}$)     &  $\mathcal{U}(-60, +60 )$ \\
$\gamma_{HIRES}$   & HIRES offset (\,m\,s$^{-1}$)     &  $\mathcal{U}(-60, +60 )$  \\
$\gamma_{EXPRESS}$   & EXPRESS offset (\,m\,s$^{-1}$)     &  $\mathcal{U}(-60, +60 )$  \\
$\gamma_{MERCATOR}$   & HERMES offset (\,m\,s$^{-1}$)     &  $\mathcal{U}(-60, +60 )$ \\
$\gamma_{PARAS}$   & PARAS offset (\,m\,s$^{-1}$)     &  $\mathcal{U}(-60, +60 )$   \\
$\sigma_{HARPS-N}$     & HARPS-N jitter (\,m\,s$^{-1}$)     &  $\mathcal{H}(5)$  \\
$\sigma_{ESPRESSO}$     & ESPRESSO jitter (\,m\,s$^{-1}$)     &  $\mathcal{H}(5)$   \\
$\sigma_{CARMENES}$     & CARMENES jitter (\,m\,s$^{-1}$)     &  $\mathcal{H}(5)$  \\
$\sigma_{NEID}$     & NEID jitter (\,m\,s$^{-1}$)     &  $\mathcal{H}(5)$   \\
$\sigma_{HIRES}$     & HIRES jitter (\,m\,s$^{-1}$)     &  $\mathcal{H}(5)$   \\
$\sigma_{EXPRESS}$     & EXPRESS jitter (\,m\,s$^{-1}$)     &  $\mathcal{H}(5)$   \\
$\sigma_{MERCATOR}$     & HERMES jitter (\,m\,s$^{-1}$)     &  $\mathcal{H}(5)$  \\
$\sigma_{PARAS}$     & PARAS jitter (\,m\,s$^{-1}$)     &  $\mathcal{H}(5)$  \\

\textit{Derived Parameters:} &       &            \\
$a/R_*$             & Scaled semi-major axis      & ...    \\
$P_{rot}$             & Stellar rotation period (days)      & ...    \\
$i$             & Transit inclination  (deg)     & ...   \\
$i_*$             & Stellar inclination  (deg)     & ...   \\
$v_{eq}\sin(i_*)$       & Projected rotational velocity  (\,km\,s$^{-1}$)     & ...    \\
$v_{eq}$       & Equatorial velocity  (\,km\,s$^{-1}$)     & ...    \\
$\psi$       & True obliquity  (deg)     & ...   \\
$\theta$ & latitude of the stellar spot (deg) & ... \\
\hline
\end{tabularx}
\footnotetext{Note: $\mathcal{N}(\mu,\sigma )$ denotes a normal distribution with $\mu$ the mean and $\sigma$ the standard deviation, $\mathcal{U}(a,b )$ denotes a uniform distribution with lower boundary $a$ and upper boundary $b$. $\mathcal{H}(\sigma )$ denotes a half-normal distribution. The limb-darkening coefficients have an uninformative prior following \citet{Kipping2013}. The impact parameter is conditioned on the value of $R_p/R_*$ }

\end{table}

\begin{table*}[t]
\setlength{\tabcolsep}{20pt}
\caption{ List of posteriors with median and $68\%$ highest density intervals}
\centering
{\begin{tabularx}{\textwidth}{@{} lccc @{}}
\hline
Parameter        &  Model with $\alpha =0$  &  Model with $\alpha =0.2$  &  Model with $\alpha$ between 0 and 1 \\
\hline
\textit{MCMC Inputs:} &    &   &              \\
Transits: &    &   &              \\
$R_*$                  &  $1.299\pm 0.002$   &   $1.299\pm 0.002$          &  $1.297\pm 0.002$     \\
$u_1$                  & $0.42 \pm 0.01$               & $0.42\pm 0.01$                 &  $0.41 \pm 0.01$   \\
$u_2$                   & $0.09_{-0.03}^{+0.02}$               & $0.09 \pm 0.03$               &   $0.11 \pm 0.02$  \\
$R_p/R_*$               & $0.0666 \pm 0.0001$    & $0.0666 \pm 0.0001$            & $0.0665 \pm 0.0001$\\
$b$                     &  $0.18 \pm 0.01$       &  $0.18 \pm 0.01$              & $0.16_{-0.02}^{+0.01}$ \\
$T_0$                   &  $58271.0738 \pm 0.0001$      &  $57186.9146\pm 0.0002 $      & $57186.9144 \pm 0.0002$\\
$T_{0,RM}$              & $59897.0199 \pm 0.0009$         &$59897.0199 \pm 0.0009$        & $59897.0196 \pm 0.0009$\\
$P_f$                   & $542.0797_{-0.0002}^{+0.0001}$ & $542.0797_{-0.0002}^{+0.0001}$ &$542.0797_{-0.0002}^{+0.0001}$\\
$\gamma_{C18}$      & $1.000003_{-0.000003}^{+0.000004}$ & $1.000003_{-0.000003}^{+0.000004}$ &$1.000003_{-0.000003}^{+0.000004}$\\
RM effect &    &   &              \\
$P_{rot}$           & $7.8 \pm 1.0$                  & $8.4 \pm 1.0$              & $8.1_{-0.9}^{+0.6}$\\

$\lambda$           & $21 \pm 8$                       & $30_{-8}^{+7}$                    & $38 \pm 3$\\
$\cos{(i_*)}$      & $0.74\pm 0.08$           & $0.68_{-0.10}^{+0.08}$            & $0.45 \pm 0.06$\\
$\alpha$            & -                                 &                             &$0.93_{-0.1}^{+0.05}$\\
$cos(\theta)$       & -                                 & $0.8_{-0.4}^{+0.2}$           &$0.94_{-0.06}^{+0.04}$\\

$\gamma_{HARPS-N}$     & $-11 \pm 1$                  & $-11\pm 1$                  & $-11\pm 1$\\
$\gamma_{ESPRESSO}$    & $-16 \pm 1$                  & $-15 \pm 1$                   & $-16 \pm 1$ \\
$\gamma_{CARMENES}$     & $-8\pm -2$       & $-8_{-2}^{+1}$                  &$-8 \pm -2$ \\
$\gamma_{NEID}$      & $-5 \pm 1$          & $-5 \pm 1$                    & $-7 \pm 1$\\
$\gamma_{HIRES}$      & $ -1.3_{-1.3}^{+1.1}$           &$0.8_{-1.4}^{+1.1}$           & $ 0.2_{-1.1}^{+1.2}$\\
$\gamma_{EXPRESS}$      & $ -3\pm 1$      & $-4 \pm 1$                    & $ -5 \pm 1$\\
$\gamma_{MERCATOR}$    & $-4 \pm 5$                 & $-4_{-5}^{+4}$                  & $-4_{-5}^{+4}$\\
$\gamma_{PARAS}$     & $ 3_{-8}^{+9}$                   & $ 3_{-10}^{+8}$                & $ 3 \pm 9$\\

$\sigma_{HARPS-N}$      & $1.13_{-1.03}^{+0.47}$        & $1.22_{-1.03}^{+0.56}$    & $1.07_{-1.00}^{+0.41}$\\
$\sigma_{ESPRESSO}$    & $1.7_{-0.4}^{+0.5}$            & $1.7_{-0.5}^{+0.4}$       & $1.18_{-0.5}^{+0.6}$\\
$\sigma_{CARMENES}$      & $3.8_{-1.1}^{+1.4}$          & $3.9_{-1.2}^{+1.3}$       &$3.6_{-1.4}^{+1.3}$\\
$\sigma_{NEID}$        & $0.6_{-0.6}^{+0.3}$            & $0.6_{-0.6}^{+0.3}$       & $0.5_{-0.5}^{+0.2}$ \\
$\sigma_{HIRES}$        & $3.2 \pm 0.4$                 & $3.2 \pm 0.4$             & $2.9 \pm 0.4$ \\
$\sigma_{EXPRESS}$      & $2.7_{-0.9}^{+0.8}$           & $2.8_{-0.8}^{+0.9}$       & $3.2 \pm 0.9$ \\
$\sigma_{MERCATOR}$     & $21 \pm 2$                    & $21\pm 2$                 & $21 \pm 2$\\
$\sigma_{PARAS}$        & $4_{-4}^{+1}$                 & $4_{-4}^{+1}$             & $4_{-4}^{+2}$\\

$a_0$                   & $-0.0002 _{-0.0089}^{+0.0111}$           & $-0.0002_{-0.0093}^{+0.0105}$       & $0.0004_{-0.0095}^{+0.0101}$ \\
$a_1$                   & $-0.01 \pm 0.09$                    & $-0.009_{-0.108}^{+0.091}$          & $-0.02_{-0.09}^{+0.10}$\\
$a_2$                   & $0.02 \pm 1.00$                 & $0.02_{-0.91}^{+1.10}$             & $0.01_{-0.93}^{+1.00}$\\

\textit{Derived Parameters:} &   &    &               \\
$a/R_*$                 &$230.04_{-0.34}^{+0.39}$             &  $230.12_{-0.37}^{+0.35}$   & $230.62_{-0.39}^{+0.38}$\\
$i$                     & $89.953 \pm 0.003 $           &  $89.954 \pm 0.003$           & $89.958 \pm 0.003 $ \\
$i_*$                   &$42_{-7}^{+6}$                 & $47 \pm 7$               &  $66 \pm 5$\\
$v_{eq}\sin(i_*)$       &$5.6 \pm 0.3$                & $6.3_{-0.4}^{+0.5}$          & $8.6_{-0.8}^{+0.9}$\\
$v_{eq}$                &$8.4_{-0.8}^{+1.4}$            & $8.6_{-0.8}^{+1.1}$               & $9.4_{-0.9}^{+1.2}$\\
$\psi$                  &$52 \pm 6$                     & $51 \pm 5$                &$45\pm3$\\
$\theta$                 & -                            & $38_{-20}^{+27}$          &$20\pm9$\\
\hline
\end{tabularx}%
}
\label{posteriors}

\label{tab_posteriors}
\end{table*}

\newpage
\clearpage
\section{ARoME analysis with macroturbulences}

\begin{figure}[h!]
  \centering
  \begin{minipage}[b]{0.45\textwidth}
    \centering
    \includegraphics[width=\textwidth]{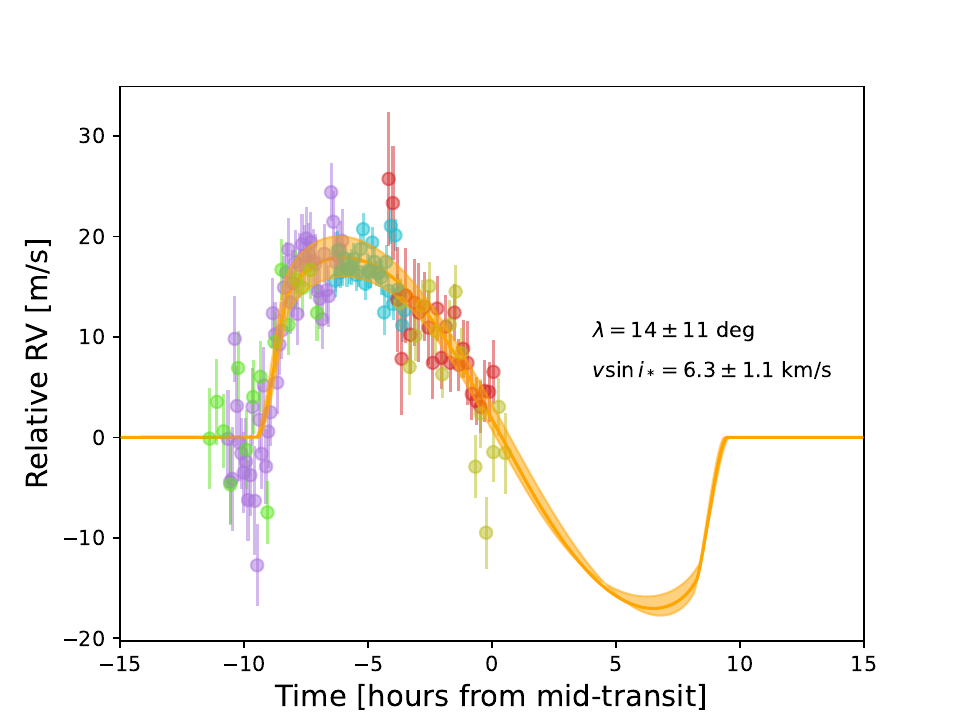}
    \caption{RM effect of the planet HIP41378 fitted with the ARoME model including macroturbulence. Only high precision RV derived using the CCF extraction method are used.}
    \label{fig_ARoME}
  \end{minipage}
  \hfill
  \begin{minipage}[b]{0.52\textwidth}
    \centering
    \begin{tabular}{lcc}
      \hline
       Parameters&Priors& Posterior\\
      \hline
       $T_0$&$\mathcal{U}(59897.010,0.1) $& $59897.030\pm 0.004$\\
       $\lambda$&$\mathcal{U}(-180,180)$ & $14 \pm 11$\\
 $v\sin(i_*)$& $\mathcal{U}(2.5,20)$&$6.4 \pm 1.0$\\
 $\zeta$ [km/s]& $\mathcal{N}(5.2,1.0)$&$3.5\pm1$\\
 $\beta_0$& $\mathcal{N}(4.2,0.8)$&$2.9\pm0.8$\\
 $\alpha_0$& fixed&4.0\\
 $P_{rot}$& $\mathcal{U}(6.0,8.5)$&$7.4\pm0.8$\\
 $i_p$& $\mathcal{N}(89.956,0.002)$&$89.95\pm0.003$\\
 $i_*$& derived&$43_{-9}^{+11}$\\
 $\psi$& derived&$49_{-11}^{+8}$ \\
 \hline
    \end{tabular}
    \vspace{32pt}
    \caption{List of posteriors with median and $68\%$ highest density intervals for the ARoME analysis.}
    \label{tab_ARoME}
  \end{minipage}
\end{figure}

\end{appendix}

\end{document}